\documentclass[journal,12pt,onecolumn,draftclsnofoot,]{IEEEtran}

\usepackage{color}
\usepackage{tikz}
\usetikzlibrary{external}
\usetikzlibrary{shapes,snakes}
\usetikzlibrary{decorations}
\usepackage{pgfplots,siunitx}
\usetikzlibrary{decorations.pathreplacing}
\usepackage{graphicx,color}
\usepackage{lipsum,adjustbox}
\usetikzlibrary{calc,positioning}
\usepackage{authblk}
\usepackage{float} % pour le placement des figure
\normalsize

\usepackage{lscape}
\usepackage{color,soul}
\usepackage[normalem]{ulem} % either use this (simple) or
\usepackage[keeplastbox]{flushend}

\graphicspath{{figures/}}

\usepackage{cite}
\usepackage{subfigure}

\usepackage{epstopdf}					% convert eps figure automatically to pdf for pdflatex
 \usepackage{booktabs}

\usepackage[cmex10]{amsmath}
\usepackage{amssymb}

\usepackage{algorithm}

\usepackage{algorithm}
\usepackage{algpseudocode}

\usepackage{xcolor}
\usepackage{mdwmath}
\usepackage{mdwtab}
\usepackage{enumitem}
\newlist{abbrv}{itemize}{1}
\setlist[abbrv,1]{label=,labelwidth=0.85in,align=parleft,itemsep=0.1\baselineskip,leftmargin=!}

\tikzset{radiation/.style={{decorate,decoration={expanding waves,angle=90,segment length=4pt}}},
         relay/.pic={
        code={\tikzset{scale=5/10}
            \draw[semithick] (0,0) -- (1,4);% left line
            \draw[semithick] (3,0) -- (2,4);% right line
            \draw[semithick] (0,0) arc (180:0:1.5 and -0.5) node[above, midway]{#1};
            \node[inner sep=4pt] (circ) at (1.5,5.5) {};
            \draw[semithick] (1.5,5.5) circle(8pt);
            \draw[semithick] (1.5,5.5cm-8pt) -- (1.5,4);
            \draw[semithick] (1.5,4) ellipse (0.5 and 0.166);
            \draw[semithick,radiation,decoration={angle=45}] (1.5cm+8pt,5.5) -- +(0:2);
            \draw[semithick,radiation,decoration={angle=45}] (1.5cm-8pt,5.5) -- +(180:2);
  }}
}
\begin{document}
\title{A survey on fiber nonlinearity compensation for \mbox{400 Gbps} and beyond optical communication systems}

%%%%%%%%%%%%%%%%%%%%%%%%%%%%%%%%%%%%%%%%%%%%%%%%%%%%
%\author{Abdelkerim.~Amari,~\IEEEmembership{Member,~IEEE},Octavia~A.~Dobre,~\IEEEmembership{Senior Member,~IEEE}, Ramachandran.~Venkatesan, Sunish.~Kumar  Philippe.~Ciblat, and Yves.~Jao{\"u}en}
\author{Abdelkerim~Amari, Octavia~A.~Dobre, Ramachandran~Venkatesan, O.~S.~Sunish~Kumar, Philippe~Ciblat, and Yves~Jaou{\"e}n

    % <-this % stops a space
%\thanks{Manuscript received April 5, 2015; revised September 13, 2015 and November 27,
%2015; accepted January 6, 2016. The editor coordinating the review process was Prof. Ekram Hossain.}     
\thanks{ Abdelkerim Amari,  Octavia A. Dobre,  Ramachandran Venkatesan, O. S. Sunish. Kumar are with the Faculty of Engineering and Applied Science, Memorial
University,  St. John's, Canada. 
Email: aamari@mun.ca.}
\thanks{Philippe~Ciblat and Yves~Jaou{\"e}n are with Telecom Paristech,
Universit{\'e} Paris-Saclay, France.}
}
\maketitle

\vspace{-0.5cm}
\begin{abstract}
Optical communication systems represent the backbone of modern communication networks. Since their deployment, different fiber technologies have been used to deal with optical fiber impairments such as dispersion-shifted fibers and dispersion-compensation fibers. In recent years, thanks to the introduction of coherent detection based systems, fiber impairments can be mitigated using digital signal processing (DSP) algorithms. Coherent systems are used in the current 100 Gbps wavelength-division multiplexing (WDM) standard technology. They allow the increase of spectral efficiency by using multi-level modulation formats, and are combined with DSP techniques to combat linear fiber distortions. In addition to linear impairments,  the next generation 400 Gbps and 1 Tbps WDM systems are also more affected by the fiber nonlinearity due to the Kerr effect. At high input powers, fiber nonlinear effects become more important and their compensation is required to improve the transmission performance. Several approaches have been proposed to deal with the fiber nonlinearity. In this paper, after a brief description of the Kerr-induced nonlinear effects, a survey on fiber nonlinearity compensation (NLC) techniques is provided. We focus on the well-known NLC techniques and discuss their performance, as well as their implementation and complexity. An extension of the inter-subcarrier nonlinear interference canceler approach is also proposed. A performance evaluation of the well-known NLC techniques and the proposed approach is provided in the context of Nyquist and super-Nyquist superchannel systems. 

%We evaluate the method by simulations in the context of Nyquist and super-Nyquist superchannel systems. A significant improvement of the transmission performance is exhibited by this extension when compared with the well-known digital back propagation and Volterra nonlinear equalizer. 
\end{abstract}

% Note that keywords are not normally used for peer review papers.
\vspace{-0.5cm}
\begin{IEEEkeywords}
Optical communication systems, nonlinear effects compensation, digital signal processing, Nyquist WDM.
\end{IEEEkeywords}

%%% list of abb

\section{INTRODUCTION}

 \IEEEPARstart{O}{ptical} communication systems have evolved since their deployment to meet the growing demand for high-speed communications. Over the past decades, the global demand for communication capacity has exponentially increased. Most of the growth has occurred in the last few years, when data started dominating the network traffic. According to Cisco Visual Networking Index (VNI) \cite{Cisco}, metro and long-haul traffic will triple between 2014 and 2019. This growth is mainly fueled by the emergence of bandwidth-hungry applications, such as cloud services and virtual reality. Furthermore, the human-centered applications, like the video games and exchange of multimedia content via smartphones, are among the most bandwidth consuming applications. In fact, in 2020, about a million minutes of video content will cross the IP network every second according to the Cisco VNI 2015-2020\cite{Cisco2}.   
\begin{figure*}[h]
  % Define the layers to draw the diagram
   \pgfdeclarelayer{background}
   \pgfdeclarelayer{foreground}
   \pgfsetlayers{background,main,foreground}
 
   % Define block styles
   \tikzstyle{interface}=[draw, fill=blue!20, text width=4em,
      text centered, minimum height=2em]
   \tikzstyle{daemon}=[
      text centered,  ]
   \tikzstyle{dots} = [above, text width=6em, text centered]
  \tikzstyle{wa} = [daemon, text width=4em, fill=red!20,
     minimum height=2em, rounded corners]
   \tikzstyle{line} = [draw]
   \tikzstyle{circle} = [draw, fill=red!20, node distance=2cm,
    minimum height=2em]
   \tikzstyle{cloud-} = [draw, ellipse,color=black!90, fill=violet!30, node distance=3cm,
    minimum height=4em, , very thick]
    \tikzstyle{cloud-1} = [draw, ellipse,color=black!90, fill=green!30, node distance=1cm,  minimum height=4em, , very thick]
    \tikzstyle{cloud-2} = [draw, ellipse,color=black!90, fill=red!30, node distance=1cm, minimum height=4em, , very thick]
    %\tikzstyle{cloud-s} = [draw, ellipse,color=black!90, fill=white!30, node distance=0.4cm, minimum height=2em, thick]
    \tikzstyle{cloud-s} = [ cloud, cloud puffs=20.7,color=black!90, fill=cyan!90, node distance=0.4cm, minimum height=0.07em, cloud ignores aspect, minimum width=2.2cm, draw]
   \begin{adjustbox}{width=\textwidth}
   \begin{tikzpicture}
      % Draw diagram elements
      %\node (wa) [wa] {$2c|\bullet|^2$};
      \path (-1,-2.0) node (n1)[cloud-] {Core optical network};
      \path (5,-2.0) node (n3)[cloud-1] {Access optical network};
      \path (2.2,-4.0) node (n2)[cloud-2] {Metro optical network};
      \path (9,-3)  pic[scale=0.5,color=blue]{relay};
      \node[inner sep=0pt] (whitehead) at (7,-4.7)
    {\includegraphics[width=.1\textwidth]{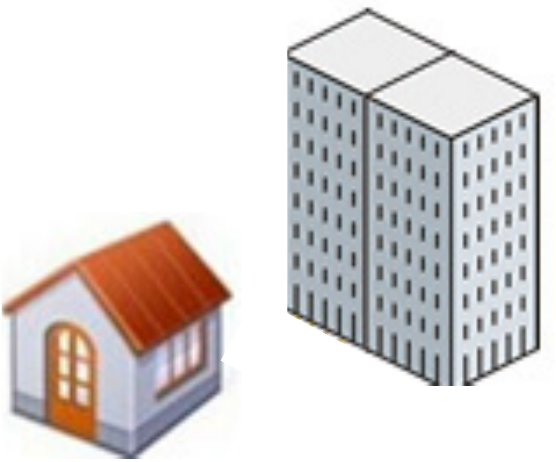}};
    % \node[inner sep=0pt] (whitehead) at (-4,-4.5)
    %{\includegraphics[width=.1\textwidth]{data.png}};
    %\path (-4,-5.8) node (gg)[daemon] {\textbf{Cloud services}};
    \path (-3.4,-3.5) node (g)[daemon] {};
    \path (7,-5.6) node (hh)[daemon] {\textbf{FTTX}};
    \path (6.9,-4.6) node (h)[daemon] {};
    \path (9,-3) node (k)[daemon] {};
    \path (9.4,-3.4) node (kk)[daemon] {\textbf{3G/4G/5G}};
   \node[cloud, cloud puffs=21.7, color=black!90, fill=cyan!30,  cloud ignores aspect, minimum width=8.2cm, minimum height=3.2cm, align=center, draw] (cloud) at (-4.2cm, -5.0cm) {\textbf{Cloud services}};
   \path (-1.8,-4.8) node (l)[cloud-s] {Emails};
    \path (-2.6,-5.7) node (l)[cloud-s] {Social};
    \path (-4.1,-4.2) node (l)[cloud-s] {Online storage};
    \path (-5.3,-5.7) node (l)[cloud-s] {Virtual Office};
    \path (-6.8,-4.85) node (l)[cloud-s] {Shared docs};
    
      % Draw arrows between elements
      \path [draw, -] (n1) -- node [above] {} (g) ;  
      \path [draw, -] (n1) -- node [above] {} (n2) ;  
      \path [draw, -] (n3) -- node [above] {} (n2) ;  
      \path [draw, -] (n3) -- node [above] {} (h) ; 
      \path [draw, -] (n3) -- node [above] {} (k) ;       
   \end{tikzpicture}
  \end{adjustbox}
\caption{Next generation of optical network. FTTX: Fiber to the (home, premises,...)}
\label{network}
\end{figure*}
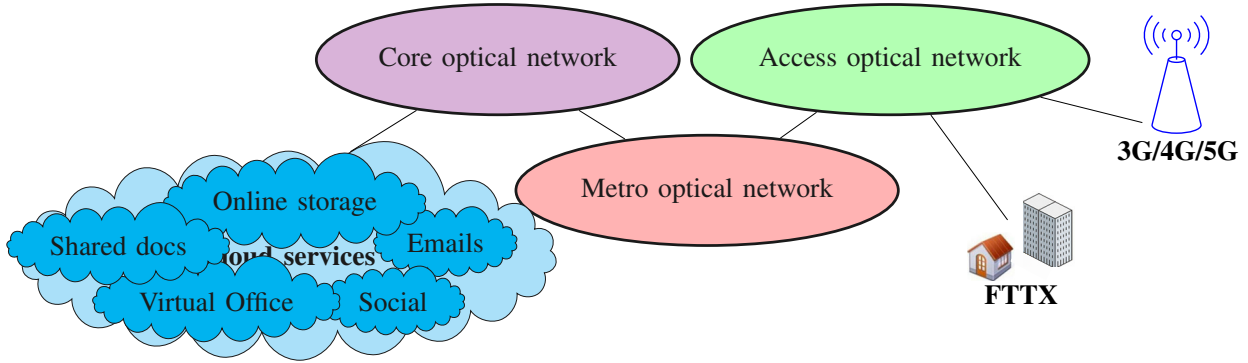
As depicted in Fig.~\ref{network}, optical communication systems represent the backbone of modern communication networks. In order to meet the increase of traffic demands, which is approaching the zettabyte threshold \cite{Cisco2}, an increase of the access network capacity, and consequently, of the metro and core network capacities is required. 
\begin{figure}[h!]
	\centering		
	\includegraphics[width=.95\linewidth]{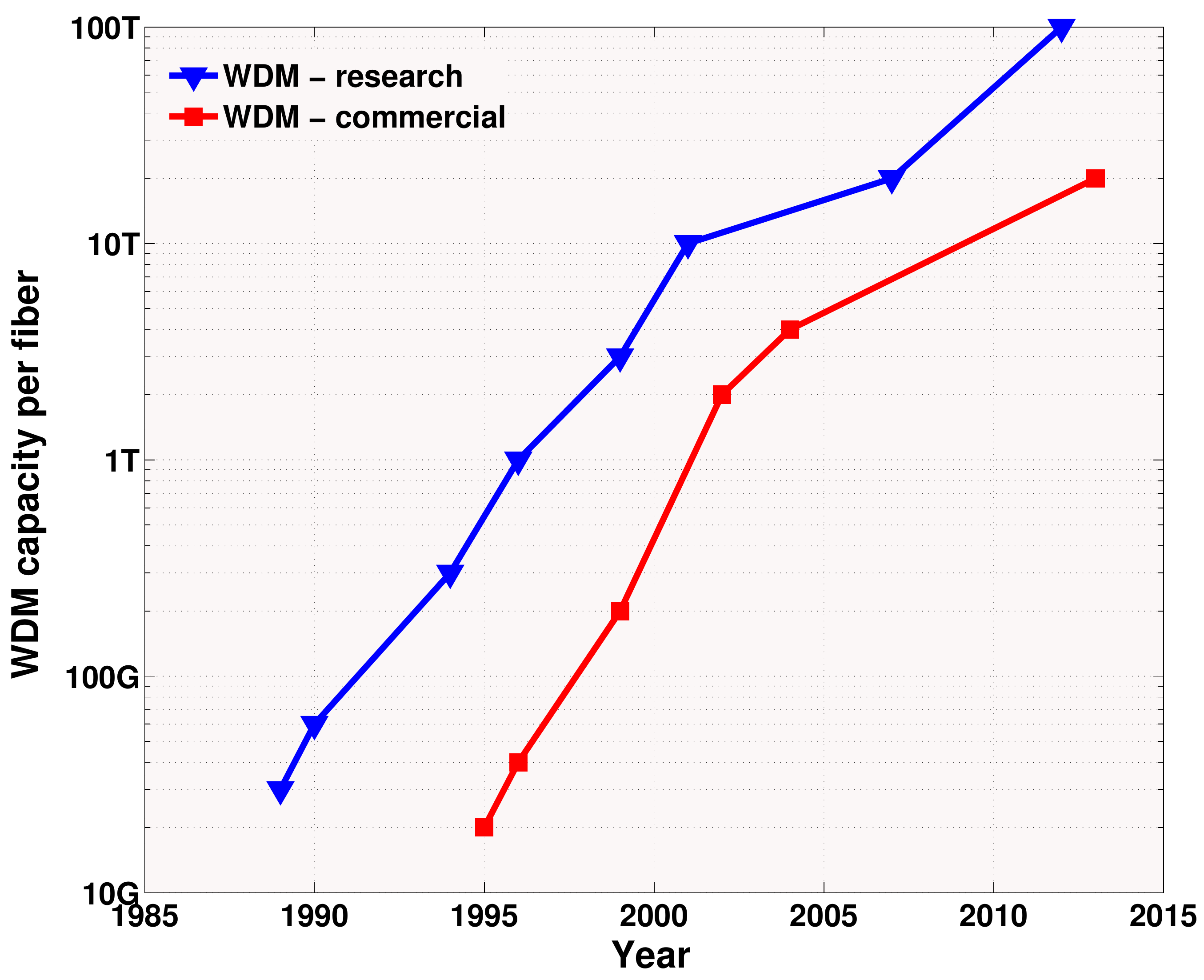}
%	\captionsetup{justification=centering}
	\caption{Evolution of the WDM capacity per fiber \cite{PJW}.}	
	\label{fig:WDM}
\end{figure}

 The deployment of the wavelength-division multiplexing (WDM) technology has been the first breakthrough that stimulated the increase of the fiber capacity. Fig.~\ref{fig:WDM} shows the evolution of the WDM capacity per fiber for both research demonstrations and commercial products \cite{PJW}. Afterwards, the re-introduction of coherent detection has revived the increase of capacity by using multi-level modulation and polarization-multiplexing transmission. Coherent systems are used in the current 100 Gb/s standard technology. To meet the continuous growth of the global demand for communication capacity, the next generation WDM communication systems are expected to operate at 400 Gb/s or 1 Tb/s rate. 
%\cite{nat} \cite{SDMM}
Different technologies are now the subject of research demonstrations to study their possible implementation in terms of complexity and costs. Space division multiplexing (SDM), such as the use of multi-core fibers and the re-introduction of multi-mode fibers, has been proposed for the next generation of WDM communication systems \cite{RJ12}--\cite{SDMexp}. In this case, the data rate can be increased according to the number of modes/cores in the multi-mode/core fibers. Some advances in the development of this technology have been achieved in recent years, especially for few-mode fibers \cite{FMF}. However, the SDM technology still faces some challenges such as the development of the optical amplifier \cite{SDMamp}, which is crucial for long-haul transmission. In addition, SDM approaches are very
expensive for the near future practical implementation because of the need to replace the already-installed single-mode fibers (SMF) by new multi-mode/core fibers. Therefore, SMF is still the technology of choice for the near future next generation
of long-haul WDM communication systems.

Researchers currently focus on increasing the transmission rate on SMF to meet the ever-increasing traffic demands. 
To achieve that, subcarrier-multiplexing, known as superchannel \cite{Bosco11}, combined with fiber nonlinearity compensation (NLC) techniques and forward error coding (FEC) \cite{FEC16}, represents the potential candidate due to its high spectral efficiency and low cost.
The main idea of the superchannel approach is to split the WDM channel into several subcarriers with smaller bandwidths and separated by small guard-band. These subcarriers are routed through optical add-drop multiplexers and wavelength selective switches as a single entity. 
The superchannel approach offers multiple advantages in comparison with single-carrier 400 Gb/s and 1 Tb/s \cite {single}. In fact, it is more flexible to the network architecture and provides higher tolerance to narrow optical filtering \cite{OIF}. In addition, it has lower requirements in terms of optical signal-to-noise ratio (OSNR) and analog-to-digital converters (ADC)/digital-to-analog converters (DAC) bandwidth \cite{OIF}. Superchannel systems also exhibit better transmission performance when compared with single-optical carrier 400 Gb/s and 1 Tb/s \cite{super}.
\begin{figure}[H]
	\centering		
	\includegraphics[width=0.95\linewidth]{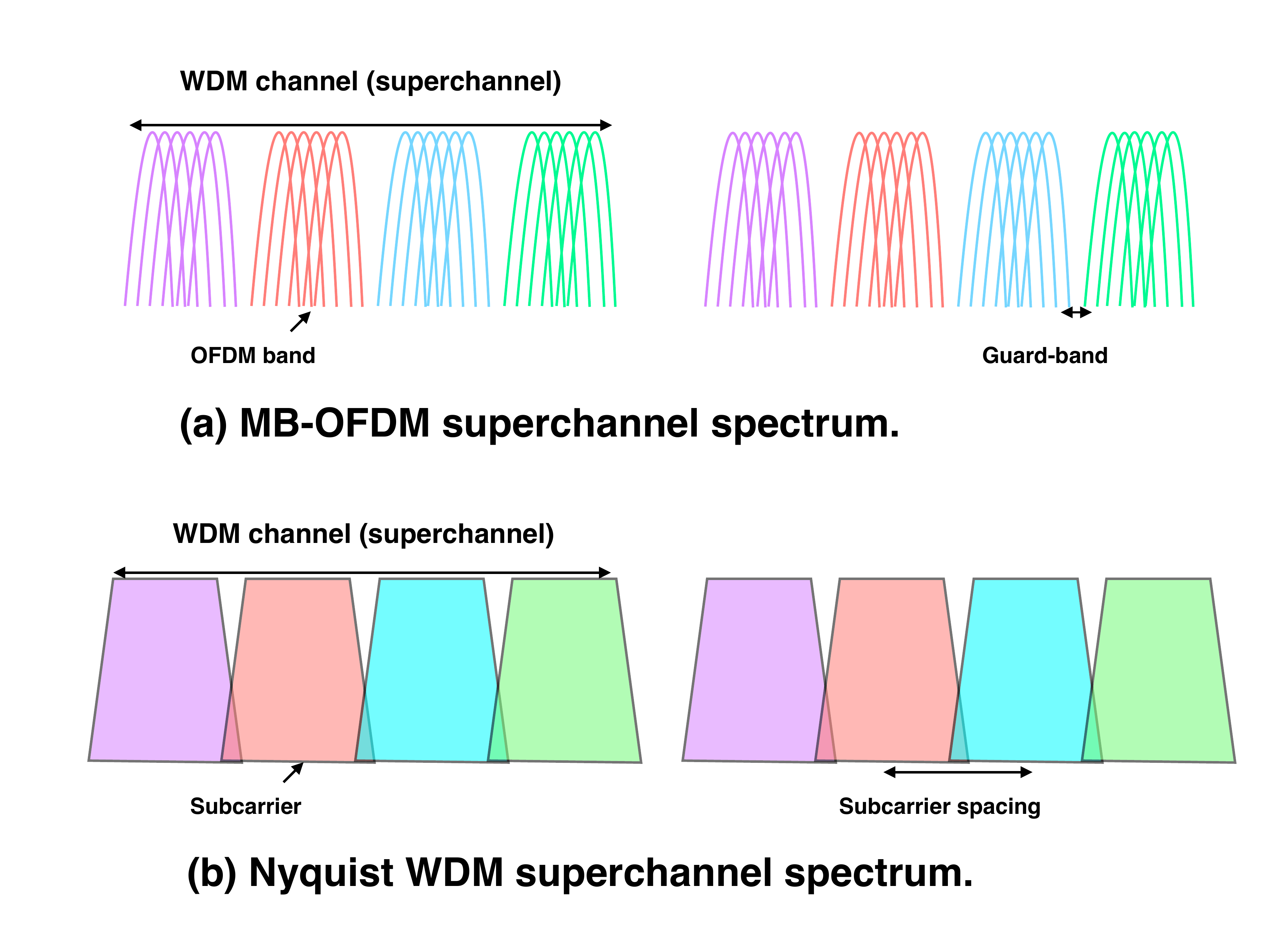}
	\caption{Superchannel transmission spectrum.}	
	\label{fig:S}
\end{figure}
%\cite{FTN1} \cite{SN1}
Two types of superchannels based on multi-band (MB) orthogonal frequency-division multiplexing (OFDM) \cite{SC12} and Nyquist WDM \cite{Bosco11} are currently investigated by the research community. The spectrum of the MB-OFDM and Nyquist WDM superchannels are shown in Fig.~\ref{fig:S}. Furthermore, a super-Nyquist\footnote{It is worth noting that super-Nyquist is also referred to as faster-than-Nyquist in the literature \cite{FTN}-- \cite{FTN2}.} WDM system, in which the subcarrier spacing is lower than the symbol rate, is also considered in the litterature to further increase the spectral efficiency \cite{JY13}--\cite{SN2}.

The superchannel transmission is a cost-effective and practical technology which can be applied in the near future. However, this kind of communication system is highly vulnerable to the fiber nonlinear effects, whose compensation is required. In fact, high-order modulations are used on each subcarrier (band in the OFDM-based superchannel) to reach the desired data rate. Such modulation formats require high OSNR, and consequently, high input power. In the Nyquist WDM system, that leads to the increase of the sensitivity to the fiber nonlinear effects, which are proportional to the instantaneous signal power. Similarly, in the MB-OFDM system, high input power leads to the increase of the peak-to-average power ratio, which results in an increase of the nonlinear distortion.
In addition, the use of smaller guard bands in superchannel systems results in substantial nonlinear inter-subcarrier interference, which significantly decreases the performance.
 
Several nonlinearity compensation (NLC) techniques have been proposed in the last decade to deal with the nonlinear effects. These techniques are applied in digital or optical domains. It is worth mentioning that there is no detailed survey of the NLC techniques in the literature. A brief description of the available techniques with a focus on the commercial application and complexity is provided in \cite{DR16}, an overview of a few NLC techniques applied in the OFDM systems is given in \cite{SNLC15}, and a description of the potential techniques to maximize the fiber capacity is provided in \cite{Cap3}. In \cite{winzer}, a recent work focusing on the nonlinear interference mitigation techniques in different practical transmission scenarios is presented. 

 In this survey paper, we discuss the state-of-the-art of the NLC techniques for both Nyquist and OFDM systems. We focus on the well-known NLC approaches, such as digital back-propagation (DBP), Volterra based nonlinear equalizer (VLNE), phase conjugation (PC) technique and perturbation-based NLC. We present a detailed theoretical description of these techniques, along with their implementation, advantages and drawbacks. An overview of other NLC techniques is given as well. Furthermore, we provide a general comparison between the main NLC techniques in terms of performance and complexity. We also extend and generalize the inter-subcarrier nonlinear interference canceler (INIC) based on Volterra series (INIC-VS) \cite{AA16} that can be used for any NLC technique, and in particular DBP.
Moreover, in the context of Nyquist and super-Nyquist superchannel systems, the performance evaluation of the well known NLC techniques, such as DBP and VLNE in addition to the proposed INIC based on DBP (INIC-DBP) is performed. We also compare the complexity of implementation of these techniques, and then a trade-off between complexity and performance is discussed.

The paper is organized as follows. In Section~\ref{section_1}, we present a brief overview of nonlinear impairments in the optical link. In Section~\ref{section_2}, we focus on the NLC techniques. We describe the well known NLC techniques, such as DBP, VLNE, PC and perturbation-based NLC, and provide an overview of other NLC approaches. This section additionally includes the description of the proposed INIC approach.
Section~\ref{section_3} is dedicated to the comparison of the main NLC techniques, along with the proposed one, in terms of performance and complexity. Finally, in Section~\ref{section_4}, we conclude the paper by giving the lessons to be learned related to the NLC techniques. 
   
\section{Overview of nonlinear impairments in the optical link}\label{section_1}
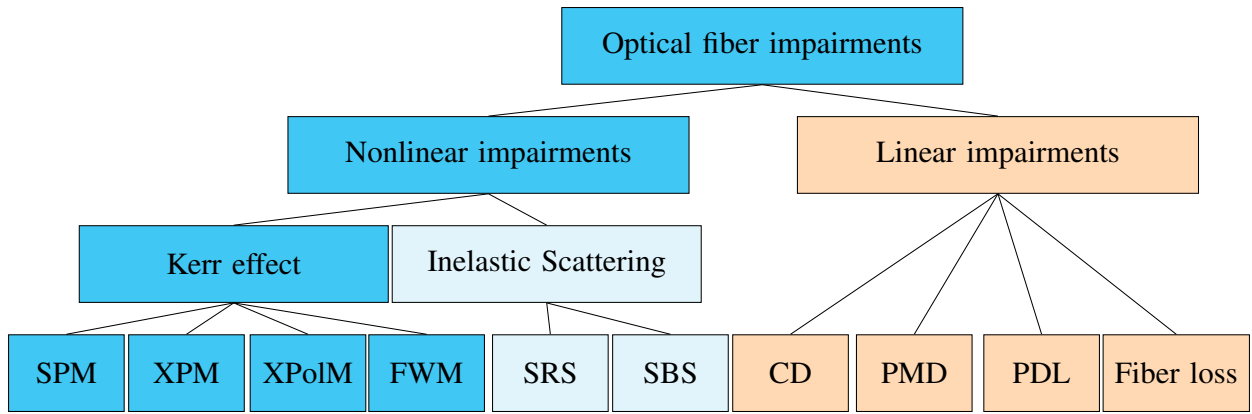
\begin{figure*}[!ht]
  % Define the layers to draw the diagram
   \pgfdeclarelayer{background}
   \pgfdeclarelayer{foreground}
   \pgfsetlayers{background,main,foreground}
 
   % Define block styles
       \tikzstyle{itnn}=[draw  , fill=cyan!60, text width=12.4em,
      text centered, minimum height=2.5em]
      \tikzstyle{itnn-}=[draw  , fill=cyan!60, text width=9.4em,
      text centered, minimum height=2.5em]
      \tikzstyle{itn}=[draw  , fill=cyan!60, text width=3.1em,
      text centered, minimum height=2.5em]
      \tikzstyle{itnn--}=[draw  , fill=cyan!10, text width=9.4em,
      text centered, minimum height=2.5em]
      \tikzstyle{it}=[draw  , fill=cyan!10, text width=3.1em,
      text centered, minimum height=2.5em]
      
      \tikzstyle{tx}=[draw, fill=orange!30, text width=12.4em,
      text centered, minimum height=2.5em]
      \tikzstyle{txx}=[draw, fill=orange!30, text width=3.1em,
      text centered, minimum height=2.5em]
      \tikzstyle{txxx}=[draw, fill=orange!30, text width=4.1em,
      text centered, minimum height=2.5em]
     \begin{adjustbox}{width=\textwidth}
   \begin{tikzpicture}
       \node (wa)[itnn] { Optical fiber impairments};
  
   \path (wa.west)+(6,-1.5) node (wl)[tx] { Linear impairments}; 
   \path (wa.west)+(-1,-1.5) node (wn)[itnn] { Nonlinear impairments}; 
   \path (wa.west)+(3.15,-4.5) node (w1)[txx] {CD}; 
   \path (wa.west)+(4.85,-4.5) node (w2)[txx] {PMD}; 
   \path (wa.west)+(6.6,-4.5) node (w3)[txx] {PDL};
   \path (wa.west)+(8.45,-4.5) node (w4)[txxx] {Fiber loss};  
   \path (wa.west)+(-4.5,-3.0)node (war1)[itnn-] { Kerr effect};
   \path (wa.west)+ (-0.2,-3.0)node (war2)[itnn--] { Inelastic Scattering};
   \path (wa.west)+(-5.15,-4.5)node (p1)[itn] {XPM};
   \path (wa.west)+(-3.48,-4.5)node (p2)[itn] {XPolM};
   \path (wa.west)+(-1.85,-4.5)node (p3)[itn] {FWM};
   \path (wa.west)+(-6.8,-4.5)node (p4)[itn] {SPM};
   \path (wa.west)+(-0.15,-4.5)node (s1)[it] {SRS};
   \path (wa.west)+(1.5,-4.5)node (s2)[it] {SBS};

 \path [draw, -] (wa.south) -- node [above] {} (wl.north) ;
  \path [draw, -] (wa.south) -- node [above] {} (wn.north) ;
  \path [draw, -] (wn.south) -- node [above] {} (war1.north) ;
  \path [draw, -] (wn.south) -- node [above] {} (war2.north) ;
  \path [draw, -] (wl.south) -- node [above] {} (w1.north) ;
  \path [draw, -] (wl.south) -- node [above] {} (w2.north) ;
  \path [draw, -] (wl.south) -- node [above] {} (w3.north) ;
  \path [draw, -] (wl.south) -- node [above] {} (w4.north) ;
  \path [draw, -] (war1.south) -- node [above] {} (p4.north) ;
  \path [draw, -] (war1.south) -- node [above] {} (p3.north) ;
  \path [draw, -] (war1.south) -- node [above] {} (p2.north) ;
  \path [draw, -] (war1.south) -- node [above] {} (p1.north) ;
  \path [draw, -] (war2.south) -- node [above] {} (s1.north) ;
  \path [draw, -] (war2.south) -- node [above] {} (s2.north) ;
   \end{tikzpicture}
   \end{adjustbox}
\caption{Optical fiber impairments. SPM: Self-phase modulation, XPM: Cross-phase modulation, XPolM: Cross-polarization modulation, FWM: Four wave mixing, SBS: Stimulated Brillouin scattering, SRS: Stimulated Raman scattering, CD: Chromatic dispersion, PMD: Polarization mode dispersion, PDL: Polarization dependent loss.}
 \label{NL-}
\end{figure*}
Optical communication over SMF suffers from several limitations. 
The diagram listing the different types of optical fiber impairments is depicted in Fig. \ref{NL-}. In addition to linear impairments, which include: chromatic dispersion (CD) \cite{CD16}, polarization mode dispersion (PMD)  \cite{AO00}, polarization dependent loss (PDL) \cite{PDL14} and fiber transmission loss \cite{loss}, nonlinear effects become a serious performance limitation at high bit rate transmissions. The optical link is a nonlinear medium due to the Kerr effect, which arises from the dependence of the optical fiber refractive index on the intensity of the transmitted signal. This effect induces different types of nonlinearity depending on the optical signal power and channel spacing (in case of multi-channel transmission), such as self-phase modulation (SPM), cross-phase modulation (XPM), four wave mixing (FWM) and cross-polarization modulation (XPolM).

Nonlinear effects can be also caused by inelastic scattering like the stimulated Brillouin scattering (SBS) and stimulated Raman scattering (SRS). 
SBS and SRS are inelastic processes in which part of the optical wave power is absorbed by the optical medium. These effects can be neglected because they manifest only at input powers higher than the typical values used in optical communication systems\cite{AG}.  

In the following, a brief description of the Kerr-induced nonlinear effects is given.
\subsection{Self-phase modulation (SPM)}
SPM consists in the signal phase change due to the interactions between the propagating signal and optical fiber. In fact, the variation of signal intensity during the propagation inside the fiber induces the variations of the refractive index, which leads to the modification of the signal phase. Thus, the nonlinear phase variation is self-induced and the related phenomenon is referred to as SPM. This causes a frequency shift, known as frequency chirping \cite{Rev11}, which interacts with the dispersion in the optical fiber and results in spectral broadening of the optical pulse \cite{SPM}. The pulse broadening increases in transmission systems with high input power because the chirping effect is proportional to the injected power.
\subsection{Cross-phase modulation (XPM)}
The communication systems are currently not limited to single-channel systems. Multi-channel transmission used in WDM systems and subcarrier multiplexing used in superchannel approaches for the next generation systems generate another type of nonlinear phase modulation, called XPM.
In this case, the fiber refractive index depends not only on the intensity of the considered optical signal but also on the intensity of other co-propagating signals \cite{XPM}. 
As a result, the nonlinear phase shift of a channel with wavelength $\lambda_j$ depends on its power $P_j$ and also on the power of other co-propagating channels $P_i,i\neq j$. As SPM, XPM reduces the transmission performance by chirping frequency and pulse overlapping. 
The XPM effect is inversely proportional to the channel spacing and increases with the number of channels or subcarriers in the context of superchannel transmission.
\subsection{Four wave mixing (FWM)}
Unlike SPM and XPM, which result in nonlinear phase shift in the optical field, the FWM process causes an energy transfer between co-propagating channels. This leads to power depletion, which degrades the performance\cite{FWM}. In addition, FWM yields inter-channel crosstalk if the generated signal falls into other co-propagating channels. This results in significant system performance degradation due to crosstalk among channels.
FWM depends on the fiber dispersion and channel spacing. As the fiber dispersion varies with the wavelength, the FWM-generated signal has a different velocity from that of the original signal. Thus, increasing the fiber dispersion limits the interactions between signals and reduces the power transfer to the newly generated signals. Increasing the channel spacing decreases the FWM effect as well. In fact, if the channel spacing is large, the FWM effect is relatively weak because the two signals walk off from each other quickly. However, FWM is more significant when the channel spacing is narrow.
\subsection{Cross-polarization modulation (XPolM)}
Polarization division multiplexing is adopted today in optical communication systems due to its improved spectral efficiency. It consists of transmitting the signal in both orthogonal states of polarization (SOPs) of the wavelength. In multi-wavelength transmission system, XPolM occurs when the SOP of a transmitted channel depends on the SOPs of other co-propagating channels which have random propagation inside the optical fiber because of PMD.
XPolM results in the depolarization of the transmitted signal, which causes fading and channel crosstalk for dual-polarization systems. XpolM can dominate the XPM effect and can be approximated as additive Gaussian noise \cite{MW10E}. 

\vspace{-0.3cm}
\subsection*{Discussion:}
The Kerr-induced nonlinear effects can be intra-channel/subcarrier (in case of superchannel transmission) nonlinear effects like SPM or inter-channel/subcarrier  nonlinear interference such as: XPM, XPolM and FWM. 
Table~\ref{tab:tab} summarizes the variation of the Kerr-induced fiber nonlinearity as a function of the bit rate and channel/subcarrier spacing. The next generation of long-haul WDM communication systems will operate at higher bit rates. Consequently, SPM, XPM and XPolM will increase, which leads to a strong reduction of the transmission performance. On the other hand, in superchannel approaches, which are adopted for the next generation systems, a smaller guard band is inserted between subcarriers. Thus, nonlinear effects such as XPM, XPolM and FWM will also increase. Moreover, in the context of super-Nyquist WDM transmission, which allows an overlap between the subcarriers, these effects became stronger and significantly reduce the transmission performance.  
\begin{table}[h!]
\centering
\caption{Fiber nonlinearity versus bit rate and channel spacing.}
 \label{tab:tab}
\begin{tabular}[width=30mm]{|c|c|c|c|c|}
       \hline  \ Type  & SPM & XPM & XPolM & FWM \\
      \hline  Bit rate $\nearrow$ & $\nearrow$ & $\nearrow$  & $\nearrow$ & no effect \\
     \hline  Channel spacing $\searrow$ & no effect  & $\nearrow$  & $\nearrow$ & $\nearrow$  \\
\hline
\end{tabular}
\end{table}

Note that other classifications of the fiber nonlinearity exist, such as the non-linear interference taxonomy proposed in \cite{PP12}. In this case, the nonlinear effects manifest as additive Gaussian noise; this is unlike the classical taxonomy, in which the nonlinear effects have different physical qualitative contributions.

\section{Fiber nonlinearity compensation techniques}\label{section_2}
Nonlinear effects mitigation is a hot research topic for increasing the fiber capacity without loss in system performance. Actually, NLC represents a key technology and a cost-effective approach to increase the data rate, being adopted for the next generation WDM systems.

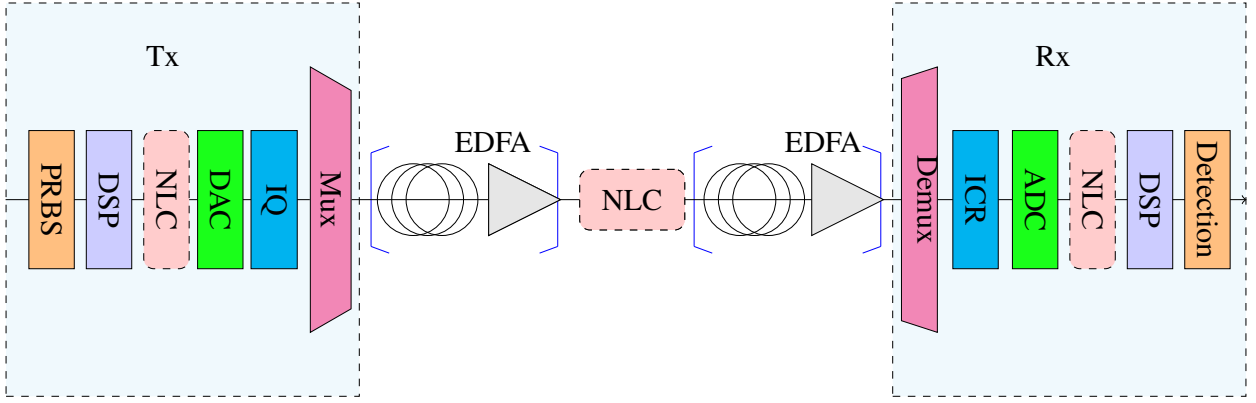
\begin{figure*}[!htp]
  % Define the layers to draw the diagram
   \pgfdeclarelayer{background}
   \pgfdeclarelayer{foreground}
   \pgfsetlayers{background,main,foreground}
 
   % Define block styles
   \tikzstyle{interface}=[draw, fill=blue!20, text width=4em,
      text centered, minimum height=2em]
      \tikzstyle{inter}=[draw  ,rotate=-90 , fill=blue!20, text width=3.9em,
      text centered, minimum height=1.5em]
       \tikzstyle{int}=[draw  ,rotate=-90 , fill=orange!50, text width=3.9em,
      text centered, minimum height=1.5em]
       \tikzstyle{itn}=[draw  ,rotate=-90 , fill=green!90, text width=3.9em,
      text centered, minimum height=1.5em]
       \tikzstyle{itnn}=[draw  ,rotate=-90 , fill=cyan!90, text width=3.9em,
      text centered, minimum height=1.5em]
      \tikzstyle{tx}=[draw, fill=cyan!5, text width=11em,
      text centered, minimum height=13em]
   \tikzstyle{daemon}=[
      text centered,  ]
      \tikzstyle{daemo}=[draw,dashed,
      text centered,  ]
      \tikzstyle{daemo--}=[rotate=-90,
      text centered,  ]
   \tikzstyle{dots} = [above, text width=7em, text centered]
  \tikzstyle{wa} = [daemo, text width=2.8em, fill=red!20,
     minimum height=2em, rounded corners]
     \tikzstyle{wawa} = [daemo,rotate=-90, text width=3.9em, fill=red!20,
     minimum height=1.5em, rounded corners]
   \tikzstyle{line} = [draw, -latex']
   \tikzstyle{circle} = [draw, fill=red!20, node distance=2cm,
    minimum height=2em]
   \tikzstyle{cloud} = [draw, fill=green!20, node distance=3cm,
    minimum height=2em]
    % \resizebox{0.5\textwidth}{!}{%
    \begin{adjustbox}{width=\textwidth}
   \begin{tikzpicture}%[remember picture, overlay]
       \node (wa)[wa] {NLC};
       \draw [fill=gray!20] (-1,0)--(-2,0.5)--(-2,-0.5)--(-1,0)--cycle;
      \path (wa.west)+(-1.2,0.8) node (d4)[daemon] { EDFA };
       \path (wa.west)+(3.4,0.8) node (d5)[daemon] { EDFA };

      \path (wa.east)+(-7.0,0.0) node [style=dashed](tx)[tx] {};
      \draw (-2.65,0) circle (0.5);
      \draw (-2.85,0) circle (0.5);
      \draw (-3.05,0) circle (0.5);
      
      \draw (1.5,0) circle (0.5);
      \draw (1.7,0) circle (0.5);
      \draw (1.9,0) circle (0.5);
      
      \path (wa.east)+(5.35,0.0) node [style=dashed](rx)[tx] {};
     
     \path (wa.west)+(-5.8,2.0) node (d5)[daemon] { Tx };
       \path (wa.west)+(6.6,2.0) node (d5)[daemon] { Rx };
       \path (wa.west)+(-2.9,0.8) node (w1)[daemon] {};
       \path (wa.west)+(-2.9,-0.8) node (w2)[daemon] {};
       \path (wa.west)+(-2.5,0.8) node (w3)[daemon] {};
       \path (wa.west)+(-2.5,-0.8) node (w4)[daemon] {};
      \path [draw,-,blue] (w1.south) -- node [above] {} (w3) ;
      \path [draw,-,blue] (w1) -- node [above] {} (w2) ;.
      \path [draw,-,blue] (w2.north) -- node [above] {} (w4) ;
      
      \path (wa.west)+(-0.3,0.8) node (e1)[daemon] {};
       \path (wa.west)+(-0.3,-0.8) node (e2)[daemon] {};
      \path (wa.west)+(-0.7,0.8) node (e3)[daemon] {};
       \path (wa.west)+(-0.7,-0.8) node (e4)[daemon] {};
      \path [draw,-,blue] (e1.south) -- node [above] {} (e3) ;
      \path [draw,-,blue] (e1) -- node [above] {} (e2) ;
      \path [draw,-,blue] (e2.north) -- node [above] {} (e4) ;
      
      % \path (wa.west)+(-0.3,-0.9) node (v)[daemon] { $\times \frac{N}{2}$ };
      %%%%%%%%%%%%%%%%%%%%%%%%%%%%%%%%%%%%%%%
       \path (wa.west)+(1.6,0.8) node (w1)[daemon] {};
       \path (wa.west)+(1.6,-0.8) node (w2)[daemon] {};
       \path (wa.west)+(2.0,0.8) node (w3)[daemon] {};
       \path (wa.west)+(2.0,-0.8) node (w4)[daemon] {};
      \path [draw,-,blue] (w1.south) -- node [above] {} (w3) ;
      \path [draw,-,blue] (w1) -- node [above] {} (w2) ;.
      \path [draw,-,blue] (w2.north) -- node [above] {} (w4) ;
      
      \path (wa.west)+(4.2,0.8) node (e1)[daemon] {};
       \path (wa.west)+(4.2,-0.8) node (e2)[daemon] {};
      \path (wa.west)+(3.8,0.8) node (e3)[daemon] {};
       \path (wa.west)+(3.8,-0.8) node (e4)[daemon] {};
      \path [draw,-,blue] (e1.south) -- node [above] {} (e3) ;
      \path [draw,-,blue] (e1) -- node [above] {} (e2) ;
      \path [draw,-,blue] (e2.north) -- node [above] {} (e4) ;
      
     % \path (wa.west)+(4.2,-0.9) node (v)[daemon] { $\times \frac{N}{2}$ };

     % \path (wa.west)+(4.5,0.0) node (a) [wa] {OPC};
     % \path (wa.west)+(4.5,0.6) node (d)[daemon] { $E \rightarrow E^*$ };
     
\node (mux)[draw, trapezium, rotate=-90,fill=magenta!60, minimum height=0.5cm,minimum width=3.7cm, trapezium stretches body] at (-4.2,0) {Mux};
\node (demux)[draw, trapezium,fill=magenta!60, rotate=90, minimum height=0.5cm,minimum width=3.7cm, trapezium stretches body] at (4.0,0) {};
\path (wa.west)+(4.85,0) node (d55)[daemo--] { Demux };
 \path [draw, -] (tx.west) -- node [above] {} (mux.south) ;
 \path [draw, -] (mux.north) -- node [above] {} (demux.north) ;
 \path [draw, ->] (demux.south) -- node [above] {} (rx.east) ;
 \path (wa.west)+(7.15,0.0) node (warx)[wawa] { NLC };   
\path (wa.west)+(7.95,0.0) node (inx)[inter] { DSP }; 
\path (wa.west)+(8.75,0.0) node (ix)[int] { Detection };
\path (wa.west)+(6.35,0.0) node (iix)[itn] { ADC };  
\path (wa.west)+(5.52,0.0) node (iiix)[itnn] { ICR };  
 \path (wa.west)+(-5.75,0.0) node (wartx)[wawa] { NLC };
 \path (wa.west)+(-4.25,0.0) node (iiitx)[itnn] { IQ };  
 \path (wa.west)+(-5.0,0.0) node (iitx)[itn] { DAC };  
 \path (wa.west)+(-7.35,0.0) node (itx)[int] { PRBS };
 \path (wa.west)+(-6.55,0.0) node (intx)[inter] { DSP }; 
\node (wa)[wa] {NLC};
  \draw[fill=gray!20] (-1,0)--(-2,0.5)--(-2,-0.5)--(-1,0)--cycle;
  \draw [fill=gray!20] (3.5,0)--(2.5,0.5)--(2.5,-0.5)--(3.5,0)--cycle;
   \end{tikzpicture}%}
   \end{adjustbox}
\caption{ Transmission diagram: possible NLC locations. PRBS: Pseudo-random binary sequences, DSP: Digital signal processing, NLC: Nonlinearity compensation, DAC: Digital-to-analog converter, IQ: In-phase and quadrature modulator, Mux: Multiplexer, Demux: De-multiplexer, ICR: Intergrated coherent receiver, ADC: Analog-to-digital converter.}
\label{RXTXLINK}
\end{figure*}

NLC techniques can be implemented in optical or digital domains. The possible locations of the proposed NLC techniques in the optical transmission link  are provided in Fig.~\ref{RXTXLINK}. Some of these techniques are applied at the transmitter side, others are done in the optical link and the majority is digitally implemented at the receiver side. In fact, due to the introduction of coherent detection, digital signal processing (DSP) algorithms have been employed to combat fiber impairments and in particular nonlinear distortion. 

In the following, a description of the most attractive NLC techniques is given. Additionally, we generalize the proposed INIC approach to compensate for both intra-subcarrier nonlinear effects and inter-subcarrier interference. A brief description of other NLC techniques is also provided.  

\subsection{Digital back-propagation (DBP)}
The DBP approach was proposed to deal with the fiber nonlinearity in digital domain. This technique can be implemented either at the transmitter, as in \cite{Rev35}, or at the receiver side \cite{EI08}. DBP is based on the split-step Fourier method (SSFM)\cite{SSF}, which represents an efficient and widely used technique to solve the Manakov equation (nonlinear Schr\"odinger equation (NLSE) in case of single-polarization transmission) given by: 
\begin{align}\label{eq:mkv} 
\frac{\partial V_{x/y}}{\partial z} + j\frac{\beta_2}{2}\frac{\partial^2 V_{x/y}}{\partial t^2} + \frac{\alpha}{2}V_{x/y} = j  \gamma' (\vert V_{x}\vert^2 + \vert V_{y}\vert^2) V_{x/y}
\end{align}
where $V = [V_{x}, V_{y}]$ is the electric field envelope of the optical signal. We denote the components of the signal $V$ on polarization $\textit{x}$ and $\textit{y}$ by $V_{x}$ and $V_{y}$, respectively. The notation $x/y$ means that, due to the symmetry, polarization $\textit{x}$ can be substituted by polarization $\textit{y}$ and vice-versa. $\alpha$ is the fiber attenuation coefficient, $\beta_2$ is the second-order dispersion parameter, $\gamma$ is the nonlinear coefficient of the fiber, and  $\gamma' = \frac{8}{9} \gamma$ is the adapted nonlinear coefficient for dual-polarization systems. The Manakov equation describes the propagation of the signal in the optical link.  
The solution of the Manakov equation is known analytically only for particular cases, such as zero-dispersion transmission. Therefore, numerical solutions, such as DBP, have been proposed. The main idea of the numerical approaches is to find a solution of the inverse Manakov equation with inverse optical link parameters, and then fiber impairments like nonlinear effects and dispersion can be mitigated. 

The DBP concept consists in transmitting the received signal through a fictitious fiber with inverse parameters. The fiber link is divided into several steps with small distance, and at each step, it is modeled as a concatenation of linear and nonlinear sections. Different ways of DBP implementation have been proposed depending on the implementation order of the linear and nonlinear parts \cite{EI08},\cite{DR11}. 
Preferably, the linear compensation part is applied first because nonlinear effects are more important at high input powers, which is the case at the end of the fictitious fiber.    

The implementation of the linear compensation section is performed in frequency domain. Using the noniterative asymmetric SSFM  \cite{SSFM}, the output of the linear section, which compensates for CD, is given by 
\begin{align}
Z_{x/y}^{\textrm{CD}}(\omega,z) = V_{x/y}(\omega,z) e^{- j h \left( \frac{\alpha}{2} + \frac{\beta_2}{2}\omega^2 \right) }
\end{align}
 where $h$ is the length of each step. Mainly, this operation corresponds to the multiplication of the received signal by an exponential term. This term represents the inverse of the signal phase change due to the dispersion. 
\begin{figure}[!h]
  % Define the layers to draw the diagram
   \pgfdeclarelayer{background}
   \pgfdeclarelayer{foreground}
   \pgfsetlayers{background,main,foreground}
 
   % Define block styles
   \tikzstyle{interface}=[draw, fill=blue!20, text width=4em,
      text centered, minimum height=4em]
   \tikzstyle{daemon}=[
      text centered,  ]
   \tikzstyle{dots} = [above, text width=7em, text centered]
  \tikzstyle{wa} = [daemon, text width=4em, fill=red!20,
     minimum height=4em, rounded corners]
   \tikzstyle{line} = [draw, -latex']
   \tikzstyle{circle} = [draw, fill=red!20, node distance=2cm,
    minimum height=2em]
   \tikzstyle{cloud} = [draw, fill=green!20, node distance=3cm,
    minimum height=2em]
     \begin{flushleft}
     \resizebox{1\textwidth}{!}{%
   \begin{tikzpicture}
      % Draw diagram elements
      \node (wa) [wa] {Nonlinear section};
      \path (wa.west)+(-3.9,0.0) node (inter)[interface] {Linear section};
      \path (wa.west)+(-0.7,0.0) node (cloud_2)[cloud] {IFFT};
      \path (wa.east)+(-7.55,0.0) node (cloud_1)[cloud] {FFT};
      \path (wa.west)+(-7.2,0.0) node (d2)[daemon] { $V_{x/y}(t)$ };
      \path (wa.west)+(3.4,0.0) node (d)[daemon] { $Z_{x/y}(t)$ };
      \path (wa.west)+(-2.1,0.25) node (dcd)[daemon] { $Z_{x/y}^{\textrm{CD}}(\omega)$ };

       \path (wa.west)+(-6.25,1.6) node (w1)[daemon] {};
       \path (wa.west)+(-6.25,-1.6) node (w2)[daemon] {};
       \path (wa.west)+(-5.85,1.6) node (w3)[daemon] {};
       \path (wa.west)+(-5.85,-1.6) node (w4)[daemon] {};
      \path [draw,-,blue] (w1.south) -- node [above] {} (w3) ;
      \path [draw,-,blue] (w1) -- node [above] {} (w2) ;
      \path [draw,-,blue] (w2.north) -- node [above] {} (w4) ;
      
      \path (wa.west)+(2.3,1.6) node (e1)[daemon] {};
       \path (wa.west)+(2.3,-1.6) node (e2)[daemon] {};
      \path (wa.west)+(2.0,1.6) node (e3)[daemon] {};
       \path (wa.west)+(2.0,-1.6) node (e4)[daemon] {};
      \path [draw,-,blue] (e1.south) -- node [above] {} (e3) ;
      \path [draw,-,blue] (e1) -- node [above] {} (e2) ;
      \path [draw,-,blue] (e2.north) -- node [above] {} (e4) ;
      
       \path (wa.west)+(2.8,-1.6) node (v)[daemon] { $\times N_s$ };
      %%%%%%%%%%%%%%%%%%%%%%%%%%%%%%%%%%%%%%%
      \path [draw, ->] (d2.east) -- node [above] {} (cloud_1.west) ;
      \path [draw, ->] (cloud_1.east) -- node [above] {} (inter.west) ;
      \path [draw, ->] (inter.east) -- node [above] {} (cloud_2.west) ;
      \path [draw, ->] (cloud_2.east) -- node [above] {} (wa.west) ;
      \path [draw, ->] (wa.east) -- node [above] {} (d.west) ;

   \end{tikzpicture}}
    \end{flushleft}
\caption{DBP implementation principle. }
\label{DBP} 
\end{figure}
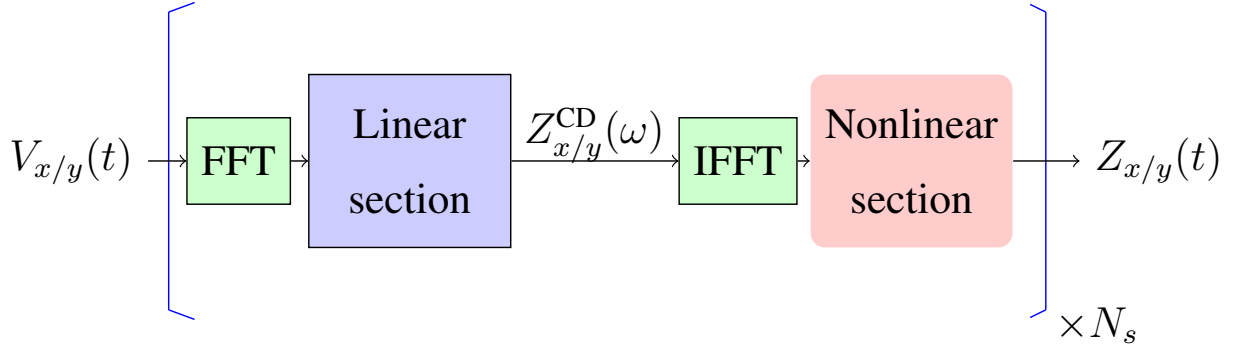
After that, the nonlinear compensation is applied in time domain to deal with  the Kerr-induced nonlinear effects. The fast Fourier transform (FFT) and inverse FFT (IFFT) are used to switch between frequency and time domains. The output of the nonlinear compensation is expressed by
\begin{align}\label{DBP-eq}
Z_{x/y}(t,z) = Z_{x/y}^{\textrm{CD}}(t,z) e^{-j \varphi \gamma' h (\vert Z_{x}^{\textrm{CD}}\vert^2 + \vert Z_{y}^{\textrm{CD}}\vert^2)} 
\end{align}
where $ 0 < \varphi < 1 $ is a real-valued optimization parameter. The exponential term introduces the phase change because of the Kerr effect. In addition to the phase change due to the self modulation of polarization $x/y$, the signal on polarization $x$ causes a nonlinear phase change of the signal on polarization $y$ and vice-versa.

The implementation of the DBP technique at the receiver side is shown in Fig.~\ref{DBP}, where $N_s$ is the number of steps. DBP can be realized either in single- or multi-step per span. It is a precise technique, which provides a high performance at small step sizes. However, it has a high computational load for real-time implementation as the number of steps per span increases. Some new approaches have been proposed to reduce the complexity of DBP based on SSFM, such as weighted DBP \cite{DRW11} and correlated DBP \cite{Rev34}; however, they are still complex for real-time implementation. DBP compensates for all deterministic impairments and is considered as the benchmark to evaluate other NLC techniques. Another DBP approach, called stochastic DBP, takes into account the noise from the optical amplifiers and is proposed to deal with the non-deterministic effects \cite{SDBP}. 
%\cite{Rev33}

In superchannel systems, the DBP performance is affected by nonlinear effects depending on the co-propagating subcarriers such as FWM, XPM and XPolM. Multi-channel DBP is proposed to combat this kind of effects \cite{MCDBP-}--\cite{MCDBP}. However, this technique, known also as total-field DBP (TF-DBP), faces the constraint of unavailability of high-speed ADC/DAC for real implementation. Furthermore, it requires a smaller step size to give better performance than the single-channel DBP \cite{TF-DBP}. A coupled-equation DBP (CE-DBP) approach has been proposed to reduce the complexity of TF-DBP \cite{TF-DBP}. CE-DBP introduces an XPM coupling term to deal with the nonlinear interference caused by adjacent subcarriers. This approach can be applied among independent receivers unlike TF-DBP, which requires the preservation of the relative phase between all subcarriers.   
Another technique based on the XPM model, called advanced DBP (A-DBP), has been also proposed for nonlinearity mitigation in superchannel systems\cite{ADBP}.

\subsection{Volterra series based nonlinear equalizer (VNLE)}
Fiber nonlinear effects can be modeled based on the Volterra series transfer function (VSFT). In fact, VSTF is a powerful tool for solving the Manakov equation (\ref{eq:mkv}) (NLSE in case of single-polarization transmission), as shown in \cite{VP97}.
After modeling the optical channel based on VSTF, the \textit{p}-th order theory developed by Schetzen \cite{MS76} is used to derive the inverse VSTF (IVSTF) kernels as a function of the VSTF ones. IVSTF kernels characterize the nonlinear equalizer which compensates for the fiber nonlinearity and CD.
Like DBP, VNLE attempts to construct the inverse of the channel. Using the \textit{p}-th order theory, up to third-order inverse Volterra operator, $K_1$ and $K_3$ can be computed from Volterra operators $H_1$ and $H_3$ as
\begin{align}
K_1 =  H_1^{-1} \\
K_3 = K_1H_3 K_1.
\end{align}
Afterwards, the IVSTF kernels are computed using the integral form of the inverse Volterra operator \cite{MS76}. Note that even order kernels are set to zero due to the isotropic property of silica, the material used for SMF.
Then, only odd-order IVSTF kernels are considered, which can be expressed based on the optical link parameters as \cite{Lui12}
\begin{eqnarray}
\label{eq:k1} k_1(\omega)\hspace{-3mm}&=\hspace{-3mm}& e^{ j\omega^2 \beta_2 NL/2} \\
\label{eq:k3} k_3(\omega_1,\omega_2,\omega-\omega_1+\omega_2)\hspace{-3mm}&=\hspace{-3mm}&\frac{jc k_1(\omega)}{4\pi^2}  \sum_{k=1}^{N} e^{j k \beta_2 \Delta\Omega L}
\end{eqnarray}
where $k_1$ and $k_3$ are the first- and third-order IVSTF kernels, respectively. 
$\omega $ is the physical optical frequency and $\omega_1$ and $\omega_2$ are the dummy variables influencing the interactions of the light waves at different
frequencies. $L$ corresponds to the span length and $\Delta \omega = (\omega_1 - \omega) (\omega_1 - \omega_2)$ is the spacing between the discrete frequencies in the sampling spectrum. The parameter $c$ is given by $c = \gamma'  L_{\textrm{eff}}$, where $L_{\textrm{eff}}$ is the effective length.

Consequently, the VNLE output can be written as a function of the received signal as
\begin{align} 
&Z_{x/y}(\omega)=  k_1 (\omega)V_{x/y}(\omega)  
 + \iint k_3(\omega_1,\omega_2,\omega-\omega_1 +\omega_2) \notag \\
&\times [V_{x}(\omega_1) V_{x}^*(\omega_2)+V_{y}(\omega_1) V_{y}^*(\omega_2)] V_{x/y} (\omega - \omega_1+\omega_2)d{\omega_1}d{\omega_2} 
\end{align}
where the superscript * stands for complex conjugation.

\begin{figure}[!h]
  % Define the layers to draw the diagram
   \pgfdeclarelayer{background}
   \pgfdeclarelayer{foreground}
   \pgfsetlayers{background,main,foreground}
 
   % Define block styles
   \tikzstyle{interface}=[draw, fill=blue!20, text width=5em,
      text centered, minimum height=1em]
   \tikzstyle{daemon}=[
      text centered,  ]
   \tikzstyle{dots} = [above, text width=6em, text centered]
  \tikzstyle{wa} = [daemon, text width=10em, fill=red!20,
     minimum height=1em, rounded corners]
   \tikzstyle{line} = [draw, -latex']
   \tikzstyle{circle} = [draw, fill=blue!20, node distance=2cm,
    minimum height=2em]
   \tikzstyle{cloud} = [draw, fill=green!20, node distance=3cm,
    minimum height=2em]
     \begin{flushleft}
      \resizebox{0.9\textwidth}{!}{%
   \begin{tikzpicture}
 
      % Draw diagram elements
      \node (wa) [wa] {Nonlinear compensation stage $1$ };
      \path (wa.west)+(1.9,1.5) node (inter)[interface] {Linear section};
      \path (wa.west)+(-2.1,1.5) node (d2)[daemon] { $V_{x/y}(\omega)$ };
      \path (wa.west)+(6.2,1.5) node (d3)[daemon] { $Z_{x/y}(\omega)$ };
      \path (wa.west)+(2.3,-1.5) node (wa1)[wa] {Nonlinear compensation stage $2$ };
      \path (wa.west)+(2.3,-3.5) node (wa2)[wa] {Nonlinear compensation stage $N$ };
      \path (wa.west)+(2.25,-2.3) node (dd)[daemon] { $\bullet$ };
      \path (wa.west)+(2.25,-2.5) node (ddd)[daemon] { $\bullet$ };
      \path (wa.west)+(2.25,-2.7) node (dddd)[daemon] { $\bullet$ };
      \path (wa.west)+(5.0,1.5) node (djyg)[daemon] {$\bigoplus$};
      \path (wa.west)+(4.95,1.5) node (djy)[daemon] {};
      \path (wa.west)+(5.0,1.47) node (dj)[daemon] {};
      \path (wa.west)+(5.0,1.6) node (d)[daemon] {};

      %%%%%%%%%%%%%%%%%%%%%%%%%%%%%%%%%%%%%%%
      \path [draw, ->] (d2.east) -- node [above] {} (inter.west) ;
      \path [draw, ->] (inter.east) -- node [above] {} (djy.west) ;
      \path [draw, ->] (d2.east) |- node [above] {} (wa.west) ;
      \path [draw, ->] (d2.east) |- node [above] {} (wa1.west) ;
      \path [draw, ->] (d2.east) |- node [above] {} (wa2.west) ;
      \path [draw, <-] (dj.south) |- node [above] {} (wa.east) ;
      \path [draw, <-] (dj) |- node [above] {} (wa1.east) ;
      \path [draw, <-] (dj) |- node [above] {} (wa2.east) ;
      \path [draw, <-] (d3.west) -- node [above] {} (inter.east) ;
   \end{tikzpicture}}
    \end{flushleft}
    \vspace{-0.5cm}
\caption{VNLE implementation.}
\label{VNLE} 
\end{figure}
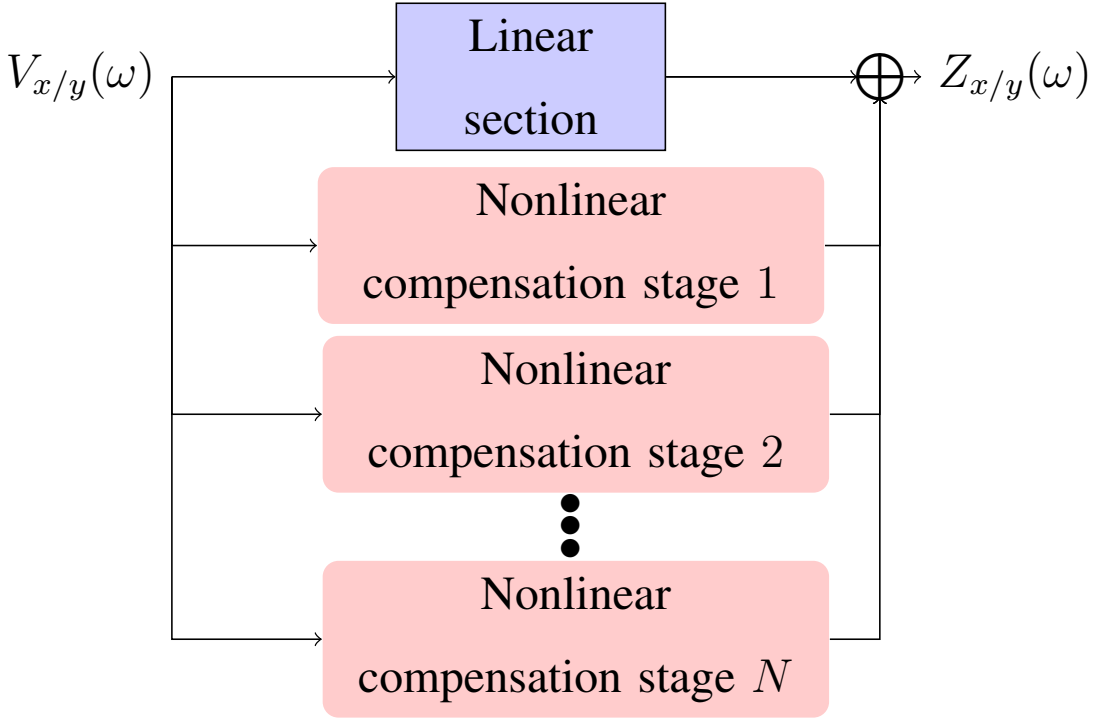

The advantage of VNLE compared to DBP is the possible parallel implementation, which reduces the computational load. %\cite{VT2}\cite{VF1}
VNLE can be processed in frequency domain \cite{JDR10}--\cite{VF2}, time domain \cite{VT1}--\cite{ZP11}, or both frequency and time domains \cite{Lui12},\cite{VTF2}. FFT and IFFT are used to pass from time domain to frequency domain and vice versa.
The principle of VNLE implementation is depicted in Fig.~\ref{VNLE}, where $N$ corresponds to the number of spans.

For each polarization, the compensation can be divided into two parts processed in parallel. The linear part consists of the CD compensation, and its output is given by
 \begin{align}
 Z_{x/y}^0(\omega) = k_1(\omega) V_{x/y}(\omega) = h_{cd}^N(\omega) V_{x/y}(\omega)
 \end{align}
where $h_{cd}(\omega) = e^{j\omega^2 \beta_2 \frac{L}{2}} \label{hcd} $ is the transfer function of CD compensation at each span. As for the DBP, the linear equalization consists of multiplying the signal by the inverse of the signal phase change due to linear impairments. 
 Concerning the nonlinear part of the compensation, it is processed in parallel for each span. The output of  each span indexed by $k$ is given by
\begin{align} 
& Z_{x/y}^k(\omega)=  \frac{j c}{4 \pi^2}\iint e^{j k \beta_2 \Delta \omega  L}V_{x/y} (\omega - \omega_1+\omega_2)  \notag \\ &\times [V_{x}(\omega_1) V_{x}^*(\omega_2)+V_{y}(\omega_1) V_{y}^*(\omega_2)] d{\omega_1}d{\omega_2}. 
\end{align}
This operation consists of multiplying the signal by an exponential term, to compensate for the phase change due to the CD, and by the total power of the signal which is the addition of the powers on polarizations $x$ and $y$. 

Finally, the output of the VNLE is obtained by combining the linear and nonlinear compensation as
 \begin{align}
 Z_{x/y} (\omega) =  Z_{x/y}^0(\omega) + \sum_{k=1}^N Z_{x/y}^k(\omega).
 \end{align}
VNLE  has shown a high performance in combating nonlinear effects for single-channel (subcarrier in superchannel systems) transmission systems, and requires about half of the DBP computational time \cite{Lui12}. New approaches have been proposed to further reduce the complexity of VNLE, such as weighted Volterra series nonlinear equalizer (W-VSNE) \cite{SBA15}. 
The common VNLE technique is based on the third-order Volterra series. A fifth-order VNLE has been also proposed \cite{Amari5},\cite{Amari2017}. While this exhibits better performance in single-channel system, it increases the complexity of implementation in comparison with the third-order case. The VNLE performance in superchannel transmission is decreased because of nonlinear interference caused by the adjacent subcarriers.

\subsection{Phase conjugation (PC)}
Different types of phase conjugation techniques have been proposed for nonlinearity mitigation, from optical phase conjugation \cite{SLJ06},\cite{OPC2} to digital phase conjugated twin waves \cite{XL13},\cite{PCTW2}.

\begin{figure}[!h]
  % Define the layers to draw the diagram
   \pgfdeclarelayer{background}
   \pgfdeclarelayer{foreground}
   \pgfsetlayers{background,main,foreground}
 
   % Define block styles
   \tikzstyle{interface}=[draw, fill=blue!20, text width=4em,
      text centered, minimum height=2em]
      \tikzstyle{tx}=[draw, fill=cyan!5, text width=3em,
      text centered, minimum height=5em]
   \tikzstyle{daemon}=[
      text centered,  ]
   \tikzstyle{dots} = [above, text width=7em, text centered]
  \tikzstyle{wa} = [daemon, text width=2.8em, fill=red!20,
     minimum height=2em, rounded corners]
   \tikzstyle{line} = [draw, -latex']
   \tikzstyle{circle} = [draw, fill=red!20, node distance=2cm,
    minimum height=2em]
   \tikzstyle{cloud} = [draw, fill=green!20, node distance=3cm,
    minimum height=2em]
     \begin{flushleft}
   \resizebox{1\textwidth}{!}{%
   \begin{tikzpicture}
      % Draw diagram elements
      %\node [circle] {Circle};  
       \draw(wa) [fill=gray!20] (1,0)--(0,0.5)--(0,-0.5)--(1,0)--cycle;
      \path (wa.west)+(2.4,0.8) node (d4)[daemon] { EDFA };
       \path (wa.west)+(8.0,0.8) node (d5)[daemon] { EDFA };
      \path (wa.east)+(-5,0.0) node [style=dashed](tx)[tx] {Tx};
      \draw (-0.9,0) circle (0.5);
      \draw (-1.1,0) circle (0.5);
      \draw (-1.3,0) circle (0.5);
      
      \draw (4.3,0) circle (0.5);
      \draw (4.5,0) circle (0.5);
      \draw (4.7,0) circle (0.5);
      
      \path (wa.east)+(5.7,0.0) node [style=dashed](rx)[tx] {Rx};

       \path (wa.west)+(0.4,0.8) node (w1)[daemon] {};
       \path (wa.west)+(0.4,-0.8) node (w2)[daemon] {};
       \path (wa.west)+(0.8,0.8) node (w3)[daemon] {};
       \path (wa.west)+(0.8,-0.8) node (w4)[daemon] {};
      \path [draw,-,blue] (w1.south) -- node [above] {} (w3) ;
      \path [draw,-,blue] (w1) -- node [above] {} (w2) ;.
      \path [draw,-,blue] (w2.north) -- node [above] {} (w4) ;
      
      \path (wa.west)+(3.3,0.8) node (e1)[daemon] {};
       \path (wa.west)+(3.3,-0.8) node (e2)[daemon] {};
      \path (wa.west)+(2.9,0.8) node (e3)[daemon] {};
       \path (wa.west)+(2.9,-0.8) node (e4)[daemon] {};
      \path [draw,-,blue] (e1.south) -- node [above] {} (e3) ;
      \path [draw,-,blue] (e1) -- node [above] {} (e2) ;
      \path [draw,-,blue] (e2.north) -- node [above] {} (e4) ;
      
       \path (wa.west)+(3.4,-0.9) node (v)[daemon] { $\times \frac{N}{2}$ };
      %%%%%%%%%%%%%%%%%%%%%%%%%%%%%%%%%%%%%%%
       \path (wa.west)+(5.9,0.8) node (w1)[daemon] {};
       \path (wa.west)+(5.9,-0.8) node (w2)[daemon] {};
       \path (wa.west)+(6.3,0.8) node (w3)[daemon] {};
       \path (wa.west)+(6.3,-0.8) node (w4)[daemon] {};
      \path [draw,-,blue] (w1.south) -- node [above] {} (w3) ;
      \path [draw,-,blue] (w1) -- node [above] {} (w2) ;.
      \path [draw,-,blue] (w2.north) -- node [above] {} (w4) ;
      
      \path (wa.west)+(8.9,0.8) node (e1)[daemon] {};
       \path (wa.west)+(8.9,-0.8) node (e2)[daemon] {};
      \path (wa.west)+(8.5,0.8) node (e3)[daemon] {};
       \path (wa.west)+(8.5,-0.8) node (e4)[daemon] {};
      \path [draw,-,blue] (e1.south) -- node [above] {} (e3) ;
      \path [draw,-,blue] (e1) -- node [above] {} (e2) ;
      \path [draw,-,blue] (e2.north) -- node [above] {} (e4) ;
      
       \path (wa.west)+(9.0,-0.9) node (v)[daemon] { $\times \frac{N}{2}$ };
     
      \path [draw, ->] (tx.east) -- node [above] {} (rx.west) ;
       \draw[fill=gray!20] (1,0)--(0,0.5)--(0,-0.5)--(1,0)--cycle;
       \draw [fill=gray!20] (6.5,0)--(5.5,0.5)--(5.5,-0.5)--(6.5,0)--cycle;
      \path (wa.west)+(4.5,0.0) node (a) [wa] {OPC};
      \path (wa.west)+(4.5,0.6) node (d)[daemon] { $V \rightarrow V^*$ };

   \end{tikzpicture}}
    \end{flushleft}
   \caption{OPC implementation.}
    \label{OPC} 
\end{figure}
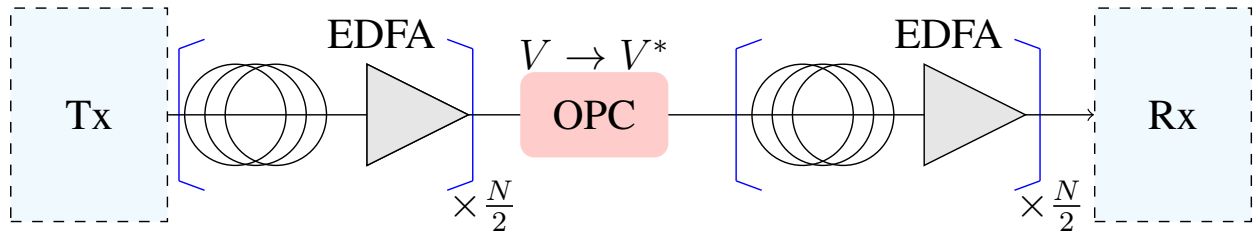

Optical phase conjugation (OPC) is implemented in optical domain and it consists in inverting the spectrum of the data signal in the middle of the transmission link, as shown in Fig.~\ref{OPC}.

The main idea of this technique is to cancel the nonlinear phase shift generated in the first segment of the link using the nonlinearity generated in the second segment of the link. However, it requires precise positioning and symmetric link design to obtain the desired performance, which significantly affects the flexibility of the optical network and make its implementation difficult. Further, OPC devices are also sensitive to the nonlinear effects, which results in additional nonlinear distortions. Recently, the multiple OPC-based nonlinear compensation technique has received significant attention and is considered as a promising approach to increase the optical systems capacity \cite{Rev32}.

Phase conjugated twin waves (PCTW) is a DSP-based approach performed at the receiver side. 
\begin{figure}[!h]
  % Define the layers to draw the diagram
   \pgfdeclarelayer{background}
   \pgfdeclarelayer{foreground}
   \pgfsetlayers{background,main,foreground}
 
   % Define block styles
   \tikzstyle{interface}=[draw, fill=blue!20, text width=4em,
      text centered, minimum height=2em]
      \tikzstyle{tx}=[draw, fill=cyan!5,, text width=6em,
      text centered, minimum height=6em]
   \tikzstyle{daemon}=[
      text centered,  ]
   \tikzstyle{dots} = [above, text width=7em, text centered]
  \tikzstyle{wa} = [daemon, text width=5em, fill=red!20,
     minimum height=7em, rounded corners]
   \tikzstyle{line} = [draw, -latex']
   \tikzstyle{circle} = [draw, fill=red!20, node distance=2cm,
    minimum height=2em]
   \tikzstyle{cloud} = [draw, fill=green!20, node distance=3cm,
    minimum height=2em]
     \begin{flushleft}
       \resizebox{1\textwidth}{!}{%
   \begin{tikzpicture}
      % Draw diagram elements
      %\node [circle] {Circle};
     % \node (wa) [wa] {Nonlinear section};
       \draw(wa) [fill=gray!20] (1,0)--(0,0.5)--(0,-0.5)--(1,0)--cycle;
      \path (wa.west)+(2.2,0.8) node (d4)[daemon] { EDFA };
      \path (wa.east)+(-6.4,0.0) node [style=dashed](tx)[tx] {};
      \path (wa.west)+(-2.0,0.0) node (inter)[interface] {PCTW};
       \path (wa.west)+(-2.0,1.5) node (d3)[daemon] { Tx };
       \path (wa.west)+(6.6,1.5) node (d3)[daemon] { Rx };
      \path (wa.west)+(-2.0,0.7) node (d2)[daemon] { $U_y = U_x^*$ };
      \draw (-0.9,0) circle (0.5);
      \draw (-1.1,0) circle (0.5);
      \draw (-1.3,0) circle (0.5);
      \path (wa.east)+(2.1,0.0) node [style=dashed](rx)[tx] {};
      \path (wa.west)+(6.6,0.0) node (interrx)[interface] {PCTW};
            \path (wa.west)+(6.6,0.7) node (d)[daemon] { $ V = [V_x ,V_y]$ };

       \path (wa.west)+(0.0,0.8) node (w1)[daemon] {};
       \path (wa.west)+(0.0,-0.8) node (w2)[daemon] {};
       \path (wa.west)+(0.4,0.8) node (w3)[daemon] {};
       \path (wa.west)+(0.4,-0.8) node (w4)[daemon] {};
      \path [draw,-,blue] (w1.south) -- node [above] {} (w3) ;
      \path [draw,-,blue] (w1) -- node [above] {} (w2) ;
      \path [draw,-,blue] (w2.north) -- node [above] {} (w4) ;
      
      \path (wa.west)+(3.6,0.8) node (e1)[daemon] {};
       \path (wa.west)+(3.6,-0.8) node (e2)[daemon] {};
      \path (wa.west)+(3.2,0.8) node (e3)[daemon] {};
       \path (wa.west)+(3.2,-0.8) node (e4)[daemon] {};
      \path [draw,-,blue] (e1.south) -- node [above] {} (e3) ;
      \path [draw,-,blue] (e1) -- node [above] {} (e2) ;
      \path [draw,-,blue] (e2.north) -- node [above] {} (e4) ;
      
       \path (wa.west)+(3.9,-0.8) node (v)[daemon] { $\times N$ };
      %%%%%%%%%%%%%%%%%%%%%%%%%%%%%%%%%%%%%%%
     
      \path [draw, ->] (inter.east) -- node [above] {} (interrx.west) ;
       \draw[fill=gray!20] (1,0)--(0,0.5)--(0,-0.5)--(1,0)--cycle;   
   \end{tikzpicture}}
    \end{flushleft}
   \caption{PCTW implementation.}
    \label{PCTW}    
\end{figure}
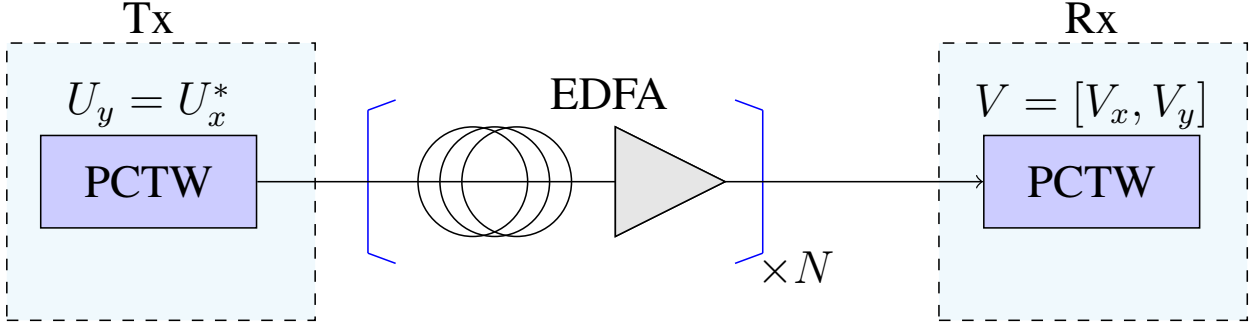
In the context of dual-polarization system, PCTW consists in transmitting the signal of interest on polarization $x$ and its conjugate on polarization $y$. The principle of the PCTW is depicted in Fig.~\ref{PCTW}.

 Assuming that the nonlinear distortions experienced by the PCTWs are anti-correlated, the first-order nonlinear phase shift can be canceled by the superposition of the two signals at the receiver side. In fact, let $\delta V_{x/y}$ be the nonlinear distortion term of the transmitted signal $U_{x/y}$. Then, the received signal can be approximated  as
\begin{align}
 V_{x/y} = U_{x/y} + \delta V_{x/y}.
\end{align} 
Knowing that the signal $ U_{y}$ is the conjugation of $ U_{x}$, the nonlinear distortion term $\delta V_y$ can be expressed in function of $\delta V_x$ as \cite{XL13}
 \begin{align}
  \delta V_y = - [\delta V_x]^*.
 \end{align}
Thus, the superposition of the received signal $V_x$ and its conjugate $V_y$ cancel the nonlinear phase shift and the transmitted signal $U_x$ can be recovered as
\begin{align}
\frac{V_x + V_y^*}{2} = U_x.
\end{align}
 This approach can be performed on the subcarrier instead of polarization in the context of coherent optical OFDM systems \cite{PCTWS}. The PCTW technique requires a pre-electrical dispersion compensation to obtain the desired performance.

PC techniques compensate for the deterministic nonlinear phase shift and also  nonlinear phase noise caused by the interaction between signal and noise. The major advantage of the PC techniques is the low complexity of implementation. In fact, PC provides an effective solution to compensate for the fiber nonlinearity because of the negligible complexity of implementation. 
On the other hand, the main drawback of PCTW is the loss of half spectral efficiency because of the transmission of the conjugate of the transmitted signal on polarization $y$. This constraint makes its implementation not efficient because of the need of full spectral efficiency. New implementations of the PCTW have been recently proposed to deal with the problem of the spectral efficiency, e.g. in \cite{YY15},\cite{sunish}. These new approaches use subcarrier coding \cite{YY15} and polarization coding \cite{sunish} in OFDM systems to double the spectral efficiency of the conventional PCTW. Another approach, called dual-PCTW, has been proposed for single carrier systems \cite{dual-pctw}. These techniques resolve the spectral efficiency issue, while they exhibit lower performance than the conventional PCTW.  

\subsection{Perturbation-based NLC} %\cite{4P}\cite{XLl14} % model \cite{SN09}
Perturbation-based approaches have been largely investigated for nonlinear effects compensation \cite{XL14}--\cite{YGa14}, as well as for modeling the optical fiber \cite{RP02}--\cite{SN14}. The perturbation-based NLC can be applied either at the transmitter side, as a predistortion, or at the receiver side. It provides an approximate numerical solution of the Manakov equation (NLSE in case of single-polarization transmission) given by (\ref{eq:mkv}). 

The main idea of the perturbation-based NLC technique is the use of the nonlinear distortion as a perturbation correction of the unperturbed solution. The unperturbed solution takes into account only the linear distortion due to dispersion and attenuation. Based on the first-order perturbation, the received field $V_{x/y}$ can be written as
\begin{align}
  V_{x/y}  = V_{x/y,0} + \gamma \delta V_{x/y}
 \end{align}
 where $V_{x/y,0}$ corresponds to the solution to linear propagation. $\delta V_{x/y}$ represents the first-order perturbation, which can be written for each polarization in frequency domain as \cite{4P}
\begin{align}
 \delta V_{x/y} (\omega, L) = h_{cd}(\omega)\int_0^L F_{x/y}(\omega,z)e^{-j\omega^2 \frac{z}{2}}d{z}
 \end{align}
where $h_{cd}$ is the transfer function of CD compensation given in \ref{hcd}, $L$ is the span length and $F_{x,y}$ is expressed as 
\begin{align}
&F_{x/y} (\omega,z) =  j\frac{8}{9}  \int e^{-j\omega t}V_{x/y,0}(t,z) [ V_{x,0}(t,z)V^*_{x,0}(t,z) +  V_{y,0}(t,z)V^*_{y,0}(t,z)]  d{t}. 
\end{align} 
 Note that the first-order perturbation coincides to the third-order Volterra series approach, as explained in \cite{RP02}. As for DBP and VNLE, the signal on polarization $x$ interacts nonlinearly with the signal on polarization $y$ and vice-versa. 
%Similarly to the VNLE, the signal is multiplied by the total power on palarizations $x$ and $y$.
When the perturbation technique is implemented as a predistortion at the transmitter side, it can be expressed for a QPSK transmission system as \cite{4P}
\begin{align}
& \delta V_{x/y} = P_0^{\frac{3}{2}} [ \sum_{m\neq0, n\neq0} A_{n,x/y} 
A^*_{m+n,x/y} A_{m,x/y} C_{m,n} +  \sum_{m\neq0, n} A_{n,y/x} A^*_{m+n,y/x} A_{m,x/y} C_{m,n} ]
\end{align} 
where $C_{m,n}$ are the nonlinear perturbation coefficients given in \cite{4P}, $A_{m/n,x/y}$ are the transmitted complex symbols and $P_0$ is the pulse peak power at the launch point. The number of perturbation coefficients depends on the pulse shape and the fiber parameters \cite{YGa14}.

The main advantage of the perturbation-based NLC techniques is the possibility of implementation on a single stage for the entire link. That significantly reduces the complexity of implementation in comparison with DBP and VNLE. It can be also implemented with one sample per symbol \cite{4P}, which reduces the requirement of the DAC/ADC speed. In addition, for relatively low spectral efficiency modulation formats, like  quadrature phase-shift-keying (QPSK) modulation, the perturbation-based NLC can be implemented without any multiplication \cite{4P}, which is not the case for higher-order modulation. An extension of multiplier-free compensator to 16-quadrature amplitude modulation (QAM) is provided in \cite{4K}, by decomposing it into two QPSK modulations.

On the other hand, the perturbation-based NLC requires a large number of perturbation terms and that affects its practical implementation. Recent research works have been proposed to reduce the number of perturbation terms \cite{YGa14},\cite{MM16}. 
\subsection{ Inter-subcarrier nonlinear interference canceler (INIC)} \label{sec-inic}
\begin{figure*}[h]
  % Define the layers to draw the diagram
   \pgfdeclarelayer{background}
   \pgfdeclarelayer{foreground}
   \pgfsetlayers{background,main,foreground}
 
   % Define block styles
   \tikzstyle{interface}=[draw, fill=blue!20, text width=5em,
      text centered, minimum height=2.5em]
      \tikzstyle{inter}=[draw   , fill=blue!20, text width=5.0em,
      text centered, minimum height=2.5em]
       \tikzstyle{int}=[draw   , fill=orange!50, text width=5em,
      text centered, minimum height=2.5em]
      \tikzstyle{interr}=[draw   , fill=blue!30, text width=6.0em,
      text centered, minimum height=2.5em]
       \tikzstyle{itn}=[draw  , fill=green!90, text width=5em,
      text centered, minimum height=2.5em]
       \tikzstyle{itnn}=[draw  , fill=cyan!90, text width=5.4em,
      text centered, minimum height=2.5em]
      \tikzstyle{tx}=[draw, fill=cyan!5, text width=35em,
      text centered, minimum height=4em]
   \tikzstyle{daemon}=[
      text centered,  ]
      \tikzstyle{daemo}=[rotate = 90,
      text centered,  ]
   \tikzstyle{dots} = [above, text width=7em, text centered]
  \tikzstyle{wa} = [daemon, text width=9em, fill=green!60,
     minimum height=2.7em, rounded corners]
     \tikzstyle{wawa} = [daemon, text width=5em, fill=red!20,
     minimum height=2.5em, rounded corners]
   \tikzstyle{line} = [draw, -latex']
   \tikzstyle{circle} = [draw, fill=red!20, node distance=2cm,
    minimum height=2em]
   \tikzstyle{cloud} = [draw, fill=green!20, node distance=3cm,
    minimum height=2em]
   \begin{adjustbox}{width=\textwidth}
   \begin{tikzpicture}
       \node (wa)[wawa] {NLC};
  \path (wa.east)+(0.3,0.0) node [style=dashed](rx)[tx] {};
    \path (wa.east)+(0.3,-2.5) node [style=dashed](rx)[tx] {};
  \path (wa.east)+(0.3,-5.0) node [style=dashed](rx)[tx] {};
   \path (wa.west)+(1.1,-5.0) node (warx)[wawa] { NLC }; 
   \node (wa)[wawa] {NLC};
 \path (wa.west)+(4.8,0.0) node (inx1)[inter] { Extra DSP }; 
\path (wa.west)+(7.4,0.0) node (ix1)[int] { Detection };
\path (wa.west)+(-2.6,0.0) node (iiix1)[itnn] { Subcarrier selection };    
\path (wa.west)+(4.8,-5.0) node (inx)[inter] { Extra DSP }; 
\path (wa.west)+(7.4,-5.0) node (ix)[int] { Detection }; 
\path (wa.west)+(-2.6,-5.0) node (iiix)[itnn] { Subcarrier selection $m_0$ };  
\path (wa.west)+(5.6,-2.5) node (inx2)[interr] { Signal \mbox{regeneration} }; 
 \path (wa.west)+(1.1,-2.5) node (warx2)[wa] { Fiber model: signal reconstruction }; 
     \path (wa.west)+(2.1,1.0) node (d5)[daemon] { Step 1 (drawn for any subcarrier $m$) };
     \path (wa.west)+(2.1,-1.5) node (d5)[daemon] { Step 2 (drawn for any subcarrier $m$) };
     \path (wa.west)+(2,-4.0) node (d5)[daemon] { Step 3 (drawn for subcarrier $m_0$) };
 \path [draw, ->] (iiix1.east) -- node [above] {} (wa.west) ;
\path [draw, ->] (wa.east) -- node [above] {} (inx1.west) ;
\path [draw, ->] (inx1.east) -- node [above] {} (ix1.west) ;
 \path [draw, ->] (iiix.east) -- node [above] {} (warx.west) ;
\path [draw, ->] (warx.east) -- node [above] {} (inx.west) ;
\path [draw, ->] (inx.east) -- node [above] {} (ix.west) ;
\path [draw, ->] (inx2.west) -- node [above] {} (warx2.east) ;
\path (wa.west)+(9.3,0.0) node (n1)[daemon] { };
\path (wa.west)+(9.9,0.0) node (n2)[daemon] { };
\path (wa.west)+(9.9,-5.0) node (n3)[daemon] { };
\path (wa.west)+(-6.2,-2.5) node (nn1)[daemon] { };
\path (wa.west)+(-5.3,-2.5) node (nn2)[daemon] { };
\path (wa.west)+(-4.5,-2.5) node (nn3)[daemon] { };
\path (wa.west)+(-5.9,-5) node (nnn1)[daemon] { };
\path (wa.west)+(-4.6,-5.0) node (nnn1)[daemon] {$\bigoplus$};
\path (wa.west)+(-5.4,-5.0) node (nnn1)[daemon] {$\bigoplus$};
\path (wa.west)+(-5.7,-4.75) node (nnn1)[daemon] {$-$};
\path (wa.west)+(-4.9,-4.75) node (nnn1)[daemon] {$-$};

\path (wa.west)+(-5.6,0.3) node (b1)[daemon] {$ V_{x/y}$};
\path (wa.west)+(-0.8,0.3) node (b1)[daemon] {$ V_{x/y,m}$};
\path (wa.west)+(2.9,0.3) node (b1)[daemon] {$ Z_{x/y,m}$};
\path (wa.west)+(9.5,0.3) node (b1)[daemon] {$ \hat{S}_{x/y,m}$};
\path (wa.west)+(3.6,-2.2) node (b1)[daemon] {$ \hat{U}_{x/y,m}$};
\path (wa.west)+(-5.3,-2.2) node (b1)[daemon] {$ W_{x/y,m}$};
\path (wa.west)+(-5.7,-3.6) node (b1)[daemo] {$ W_{x/y,m_0-1}$};
\path (wa.west)+(-5.0,-3.6) node (b1)[daemo] {$ W_{x/y,m_0+1}$};
\path (wa.west)+(-0.7,-4.7) node (b1)[daemon] {$ V_{x/y,m_0}^{inic}$};
\path (wa.west)+(9.5,-4.7) node (b1)[daemon] {$ \hat{S}_{x/y,m_0}^{inic}$};

\path (wa.west)+(-5.9,-0) node (n)[daemon] { };
\path (wa.west)+(-6.2,-0) node (ns)[daemon] { };
\path [draw, ->] (n1.east) |- node [above] {} (inx2.east) ;
\path [draw, ->] (ix1.east) -- node [above] {} (n2.west) ;
\path [draw, ->] (ix.east) -- node [above] {} (n3.west) ;
\path [draw, -] (warx2.west) -- node [above] {} (nn1.west) ;
 \path [draw, -] (ns.west) -- node [above] {} (iiix1.west) ;
 \path [draw, ->] (n.west)|- node [above] {} (iiix.west) ;
\path [draw, ->] (nn2.west)|- node [above] {} (iiix.west) ;
\path [draw, ->] (nn3.west)|- node [above] {} (iiix.west) ;
   \end{tikzpicture} 
   \end{adjustbox}
\caption{ INIC implementation.}
\label{INIC}
\end{figure*}
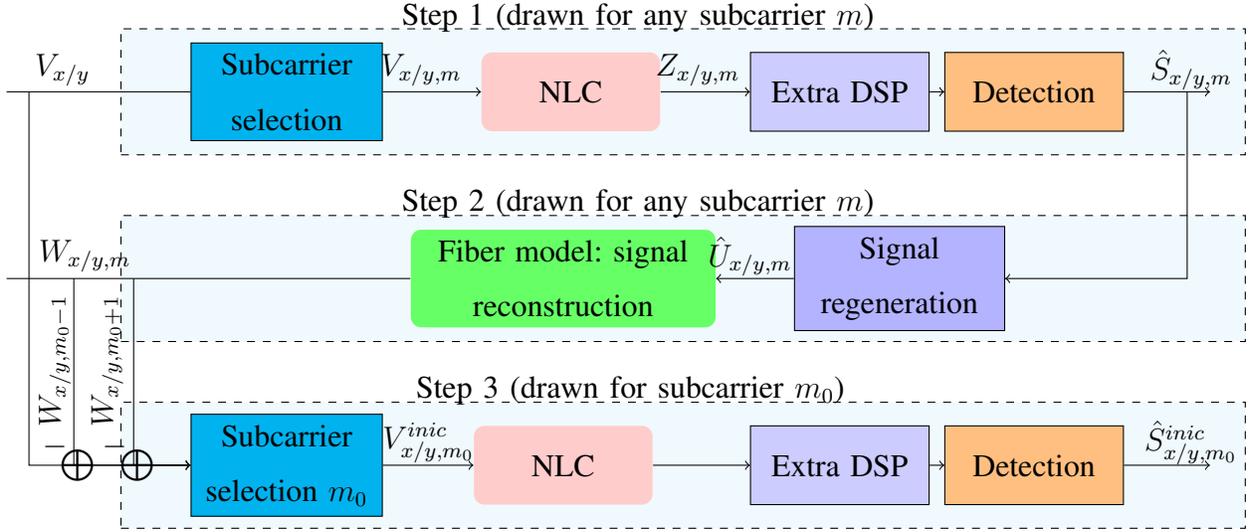

In the context of Nyquist and super-Nyquist WDM superchannel, the interference caused by the adjacent subcarriers severely affects the transmission performance. In addition to SPM, which is the intra-subcarrier nonlinear effect, nonlinear interference caused by XPM, XpolM and FWM significantly increase, and mitigation of these impairments is required. This interference reduces the performance of the classic NLC techniques, such as the single channel/subcarrier DBP and VNLE. Then, to deal with nonlinear and linear interference, the INIC approach based on the Volterra series (INIC-VS) was proposed  in \cite{AA16} (INIC(3,3) in \cite{AA16}). This technique is based on the decision feedback equalizer (DFE) \cite{DDF78}. The main idea is to make use of the prior knowledge of the detected adjacent subcarriers to cancel the interference on the subcarrier of interest. It consists in detecting the adjacent subcarriers, regenerating them using the Volterra series fiber model, and finally removing them from the subcarrier of interest. 
  
Here, we generalize the INIC approach such that it can be implemented along with other NLC techniques. In fact, as shown in Fig.~\ref{INIC}, the implementation of the INIC approach can be divided into three steps:
\begin{itemize}
\item In the first step, the received signal is passed through a subcarrier selection because the receiver proceeds subcarrier per subcarrier. Then, a nonlinear equalizer is applied to deal with intra-subcarrier nonlinear effects. This nonlinear equalizer can be a VNLE, as in \cite{AA16}, a DBP, a perturbation-based NLC, PCTW or any other NLC techniques. Afterwards, extra DSP is required to compensate the phase and frequency offset and deal with the PMD and residual dispersion. Finally, a threshold detector is applied for signal detection.

\item In the second step, the detected signals of the adjacent subcarriers are firstly re-modulated, and then reconstructed based on the optical fiber model. This model can be the Volterra series fiber model, digital propagation or any other fiber model. 
 
\item In the third step, if we consider $m_0$ as the subcarrier of interest, the rebuilt signals of adjacent subcarriers $W_{x/y, m_0 -1}$ and $W_{x/y, m_0 +1}$ are removed from the original received signal $V_{x/y}$. After that, a similar process is applied as in the first step. A nonlinear equalizer, which can be any type of the NLC techniques, is used to compensate for intra-subcarrier nonlinear effects. 
\end{itemize}

Hereafter, we consider INIC based on DBP (INIC-DBP) as an example and present its principle of implementation. In the INIC-DBP approach, DBP is used for intra-subcarrier nonlinear effects compensation. The output of the DBP based equalizer is given in (\ref{DBP-eq}). Concerning the recursive part, the digital propagation fiber model is used to reconstruct the regenerated detected signal $\hat{U}_{x/y,m}$ of each adjacent subcarrier. It can be determined from the DBP technique by inverting the sign of the fiber parameters ($\beta_2$, $\alpha$, $\delta$) and the gain of the EDFA amplifier. The output of the digital propagation fiber model  for each subcarrier and polarization is expressed as 
\begin{align}
W_{x/y,m}(t,z) = W_{x/y,m}^{\textrm{CD}}(t,z) e^{j \varphi \gamma' h (\vert W_{x,m}^{\textrm{CD}}\vert^2 + \vert W_{y,m}^{\textrm{CD}}\vert^2)} 
\end{align}
where the linear model $\hat{W}_{x/y,m}^{\textrm{CD}}$ is given by
\begin{align}
W_{x/y,m}^{\textrm{CD}}(\omega,z) = \hat{U}_{x/y}(\omega,z) e^{ j h \left( \frac{\alpha}{2} + \frac{\beta_2}{2}\omega^2 \right) }.
\end{align}
Finally, in the third step, the contributions of the closest adjacent subcarriers $m_0 -1$ and $m_0+1$ are subtracted from the original received signal $V_{x/y}$ and the new receiver input is given by
\begin{align}
V_{x/y}^{\textrm{inic}}(\omega) = V_{x/y}(\omega) -  W_{x/y,m_0-1}(\omega) - W_{x/y,m_0+1}(\omega).
\end{align}
After the selection of the subcarrier of interest $m_0$, a DBP-based equalizer is applied to compensate for the intra-subcarrier nonlinear effects, and then the final decision is made.

Note that, in this proposed INIC scheme, we consider only the nonlinear interference caused by the adjacent subcarriers. In fact, the received signal on subcarrier $m_0$ can be written after the subcarrier selection as:
\begin{align}
V_{x/y,m_0} = f_1(U_{m_0}) + f_2(U_{m_0},\bar{U}_{m_0}) + f_3(\bar{U}_{m_0})
\end{align}
where
$\bar{U}_{m_0} = \lbrace U_m \rbrace_{m=1, m \neq m_0}^M$, with $M$ as the number of transmitted subcarriers. In terms of nonlinearity, $f_1$ represents the intra-subcarrier nonlinear effect, while $f_2$ and $f_3$ represent the inter-subcarrier nonlinear interference.
In the proposed INIC approach, we reconstruct and then subtract only the inter-subcarrier nonlinear interference due to the term $f_3$. The inter-subcarrier nonlinear interference due to the term $f_2$ and the intra-subcarrier nonlinear effect due to $f_1$ are not taken into account. These two terms cause a causality issue due to the existence of the current symbol and precursor nonlinear interference. Some approaches to deal with nonlinear interference caused by current symbol and precursor interference have been proposed in wireless communication systems, such as the root method \cite{Root} and the precursor enhanced RAM-DFE canceler \cite{RAM-DFE}. Such methods have not yet been investigated for optical communications, to the best of our knowledge.  

In addition to intra-subcarrier nonlinear effects, the INIC approach compensates for inter-subcarrier linear and nonlinear interference, which represent major challenges in super-Nyquist and Nyquist WDM communication systems. On the other hand, the INIC-DBP technique roughly triples the complexity in comparison to the single-step per span DBP. More details about complexity are given in the next section.

\subsection{Other proposed approaches for NLC}
 Several research works have studied the DBP, VNLE, PC, and perturbation-based NLC techniques. A combination of some of them have been also investigated, e.g., the combination of DBP and OPC (spectral inversion) \cite{DBPSI},\cite{dr011}, and the combination of DBP and perturbation based NLC \cite{XL14}.
 In addition to these techniques, various NLC approaches have been proposed to compensate for fiber nonlinearity.
%\cite{NFT3}

A new approach, based on eigenvalue communication \cite{NFT0}, has been recently proposed. The main idea is to encode the transmitted information in the nonlinear Fourier transform (NFT) of the signal, due to the integrability\footnote{The integrability of the optical system means that the NLSE can be represented by a Lax pair $[L,M]$. The main point of the Lax pair is that the eigenvalues of the linear operator $L$ are independent of time\cite{NFT0}.} of the optical fiber \cite{NFT1}--\cite{NFT4}. The NFT consists of a continuous and a discrete spectrum. Some approaches use the discrete part of the spectral function, which corresponds to the soliton transmission, to modulate the signal \cite{NFT3}. Another approach, called nonlinear inverse synthesis, modulates the signal on the continuous part of the NFT spectrum \cite{NFT2}.
Nonlinear inverse synthesis exhibits similar transmission performance to DBP. It has a comparable complexity of implementation to DBP \cite{NFT2}.
NFT-based communication is not affected by all linear and nonlinear deterministic effects including intra-channel and inter-channel cross-talk. It can be considered as a promising candidate to be used for the future optical communication systems. On the other hand, NFT-based communication relies critically on the integrability of the optical channel, which can be disturbed by some effects, such as the fiber loss and hardware related distortions. In addition, it is also limited by the interaction between the signal and the noise introduced by the EDFAs. 

 Machine learning-based techniques, such as support vector machine equalization \cite{SVM}, have been investigated to deal with the fiber nonlinearity. In this approach, multiple two-class support vector machines are used to build a multi-class classifier, which consists of constellation clusters. The main idea is to use a training and testing process, respectively. The training process is to determine the distribution of the possibly noisy constellation points. Then, the testing process compares the predicted output of the support vector machine equalizer with the pre-stored transmitted symbols. The support vector machine-based classification equalizer has been considered to compensate both deterministic nonlinear effects and non-deterministic nonlinear phase noise\cite{SVM}--\cite{SVM2}. The nonlinear phase noise is caused by the interaction of the signal with the amplified spontaneous emission (ASE) noise, introduced by the optical amplifier.

%\cite{dr010} 
In the context of coherent optical OFDM system, RF-pilot tones \cite{pilot} and Wiener-Hammerstein model-based electrical equalizer \cite{JP11} have been considered to combat the fiber nonlinearity. The RF-pilot tones compensate for the XPM nonlinear effects, being inspired from the RF-pilot based phase noise compensation. The nonlinear distortions can be compensated by firstly inverting the RF-pilot phase and then multiplying it with the OFDM symbol. The Wiener-Hammerstein model-based electrical equalizer is a similar technique as the VNLE. In this approach, finite impulse response filters are deployed as linear filters and a polynomial with only odd-order terms is used as the memoryless nonlinearity \cite{JP11}. The Wiener-Hammerstein model technique has a lower complexity in comparison with the Volterra model. However, the Volterra-based nonlinear equalizer considers a memory for nonlinearity compensation, which can provide better results. 

Other NLC approaches have been also proposed in single-carrier communication systems such as: optical back-propagation \cite{OBP}, code-aided expectation-maximization algorithm \cite{CAEMA} and electronic compensation technique \cite{ECT}. Optical back-propagation is implemented on the optical link. It divides each span into several sections, and for each section, fiber Bragg gratings and highly nonlinear fibers are used to compensate for the dispersion and fiber nonlinearity, respectively. The code-aided expectation-maximization algorithm and the electronic compensation technique are both used to compensate for the nonlinear phase noise at the receiver side.
An adaptive maximum likelihood sequence detection (MLSD) algorithm has been also proposed in single carrier systems \cite{MLSD}. Such a detector is used to mitigate the nonlinear phase noise. The MLSD can be also combined with other NLC techniques such as DBP \cite{DBPMLSD}, and consequently, both deterministic and non-deterministic nonlinear effects can be compensated for.  

{In dual-polarization systems, the nonlinear cross-talk between the polarizations, known as XPolM, represents a strong limitation of the performance. A nonlinear polarization crosstalk canceller based on multiple-output multiple-input eqalization has been proposed in \cite{Rev31} to deal with such an effect.}

 \vspace{-0.15cm}
\subsection*{Discussion:}
\begin{table*}[htp!]
\centering
\caption{Fiber nonlinearity compensation techniques.} \label{tab:table2}
  \begin{adjustbox}{width=\textwidth}
\begin{tabular}[width=80mm]{|c|c|c|c|c|c|}
       \hline \ \textbf{Technique} & \textbf{Type} & \textbf{Location} & \textbf{Fiber nonlinearity compensated} & \textbf{Transmission system} & \textbf{References}  \\
       \hline    Digital back propagation (DBP) & Digital & Tx/Rx & Intra-subcarrier & Nyquist/OFDM & \cite{EI08}--\cite{ADBP} \\
       \hline    Total-field digital back propagation (TF-DBP) & Digital & Tx/Rx & Intra- and inter subcarrier & Nyquist/OFDM & \cite{MCDBP-}-- \cite{TF-DBP} \\
       \hline    Volterra-based nonlinear equalizer (VNLE)& Digital & Tx/Rx & Intra-subcarrier & Nyquist/OFDM & \cite{VP97}--\cite{Amari5} \\
       \hline    Phase conjugation (PC)& Digital/Optical & Rx/Link & Nonlinear phase & Nyquist/OFDM &\cite{SLJ06}--\cite{dual-pctw} \\
       \hline    Perturbation-based NLC & Digital & Tx/Rx & Intra-subcarrier/XPM & Nyquist/OFDM & \cite{XL14}--\cite{4K} \\
       \hline    Inter-subcarrier nonlinear interference canceler (INIC) & Digital & Rx & Intra- and inter-subcarrier & Nyquist/OFDM & \cite{AA16},\cite{Am} \\
       \hline Nonlinear Fourier transform & Digital & Tx/Rx & Intra- and inter-subcarrier & Nyquist/OFDM & \cite{NFT0}--\cite{NFT2} \\
     \hline  Wiener-Hammerstein  &  Digital & Rx & Intra-subcarrier & OFDM & \cite{JP11}  \\
     \hline    Radio frequency (RF)-pilot tones & Digital & Rx & Nonlinear phase shift & OFDM & \cite{pilot} \\
     \hline    Support vector machine & Digital & Rx & Intra-subcarrier/nonlinear phase noise & Nyquist/OFDM & \cite{SVM}--\cite{SVM2} \\ 
     \hline Optical back propagation  & Optical & Link & Nonlinear phase & Nyquist & \cite{OBP} \\ 
     \hline Code-aided expectation-maximization algorithm  & Digital & Rx & Nonlinear phase noise  & Nyquist & \cite{CAEMA} \\ 
     \hline Electronic compensation technique  & Digital & Rx & Nonlinear phase noise & Nyquist & \cite{ECT} \\ 
     \hline  {Adaptive maximum likelihood sequence detection (MLSD)}  &  {Digital} &   {Rx} &  {Nonlinear phase noise} &   {Nyquist} &  {\cite{MLSD}--\cite{DBPMLSD}} \\
      \hline  {Nonlinear polarization crosstalk canceller}  &   {Digital} &   {Rx} &   {XPolM} &  {Nyquist} &   {\cite{Rev31}} \\ 
\hline
\end{tabular}
\end{adjustbox}
\end{table*}
Various NLC techniques have been subject of research investigations in the last decade to evaluate their performance in different communication systems, as enlisted in Table \ref{tab:table2}. The table \ref{tab:table2} shows also the type and location of the NLC techniques, in addition to the type of the fiber nonlinearity which they compensate for. Note that the term nonlinear phase represents the deterministic nonlinear phase shift (SPM and XPM) as well as the non-deterministic nonlinear phase noise due to the interaction between signal and noise.

DBP compensates for all deterministic effects and provides high performance at small step size. However, it is not considered for real implementation because of the high computational load. Concerning VNLE, this technique has relatively lower complexity when compared with DBP due to parallel implementation, but its complexity is still high for commercial implementation. Furthermore, these two techniques are affected by the nonlinear interference in superchannel systems.
  
PCTW and OPC techniques have the advantage of reduced complexity. However, OPC faces the problem of flexibility because it requires a precise positioning and a symmetric link. On the other hand, the conventional PCTW technique engenders the loss of half spectral efficiency due to the transmission of signal conjugation.

Perturbation-based NLC approaches can be implemented in one step for the entire link and with one sample per symbol, which is not the case of DBP and VNLE. However, the perturbation-based NLC is still complex for implementation because it requires a large number of perturbation terms.

In the context of superchannel systems, the nonlinear interference becomes a strong limitation of performance. Some approaches such as INIC, NFT and TF-DBP deal with this kind of distortion, which leads to better performance but also higher complexity.     

\section{Results and discussions} \label{section_3}
\subsection{Simulation setup} 
In the context of Nyquist and super-Nyquist WDM transmission, we evaluate the performance of the DBP, TF-DBP and VNLE approches, along with that of the proposed INIC-DBP and the INIC-VS. To quantify the effect of nonlinear interference, we introduce the inter-subcarrier linear interference canceler (ILIC) (INIC(1,1) in \cite{AA16}). ILIC compensates for only the linear interference and CD without taking into account the nonlinear effects.

To do this, we generate a dual-polarization 16QAM modulated Nyquist-WDM superchannel with 4 subcarriers. The bit rate is $448$ Gb/s and the symbol rate per subcarrier and per polarization is $R=14$ Gbaud. The transmission line consists of multi-span standard SMF with $\alpha=0.2~\mathrm{dB.km^{-1}}$, $D=17 ~\mathrm{ps.nm^{-1}.km^{-1}}$, and \mbox{$\gamma=1.4~ \mathrm{W^{-1}.km^{-1}}$}. The polarization mode dispersion (PMD) is $0.1~ \mathrm{ps.km^{-1/2}}$. An EDFA with a $5.5$ dB noise figure and  $20$ dB gain is used at each span of $100$ km. Table.~\ref{tab:tab3} summarizes the link parameters used for simulations.
\begin{table}[h]
\centering
\caption{Link parameters.} \label{tab:tab3}
\begin{tabular}[width=30mm]{|c|c|}
\hline  Attenuation coefficient ($\alpha$) & $0.2~dB.km^{-1}$ \\
\hline  Dispersion parameter (D) & $17 ~ps.nm^{-1}.km^{-1}$ \\
\hline  Nonlinear coefficient ($\gamma$)& $1.4~W^{-1}.km^{-1}$\\
\hline Polarization mode dispersion (PMD)& $0.1~ ps.km^{-1/2}$ \\
\hline  EDFA noise figure  & $5.5$ dB \\
\hline  EDFA gain & $20$ dB \\
\hline  Span length $L$  & $100$ km\\
\hline
\end{tabular}
\end{table}

A root-raised cosine (RRC) filter with a roll-off factor $\rho$ is used to shape the spectrum of the subcarriers. Note that the ADC works at twice the symbol rate. The transmission parameters are given in Table.~\ref{tab:table1}. 
\begin{table}[h]
\centering
\caption{Transmission parameters.} \label{tab:table1}
\begin{tabular}[width=30mm]{|c|c|}
\hline  Subcarrier number ($M$)  & $4$ \\
\hline  Bit rate & $448$ Gbps \\
\hline  Symbol rate ($R$)  & $14$ GBd\\
\hline  Modulation  & $16$QAM\\
\hline  RRC roll-off factor ($\rho$) & $0.1$ or $0.01$\\ 
\hline  ADC samples per symbol & $2$ \\ 
\hline
\end{tabular}
\end{table}

In addition to CD and nonlinear compensation, an adaptive constant modulus algorithm is applied to handle the PMD and the residual dispersion \cite{CMA}. The constant phase estimation is carried out by applying the Viterbi-Viterbi algorithm \cite{CPE}.

\subsection{Performance evaluation} 
The performance of the NLC techniques are shown in terms of the Q factor and subcarrier spacing factor. The Q factor is related to the bit-error rate (BER) as \cite{BERQ}
\begin{align}
Q = 20\log (\sqrt{2}\textrm{erfc}^{-1}(2\textrm{BER})).
\vspace{10pt}
\end{align}
The subcarrier spacing factor $\Delta$ is defined as the ratio between the subcarrier spacing $\Delta f$ and the symbol rate $R$, i.e., $\Delta = \Delta f/R$.

All results concern the central subcarriers, as they are more disturbed by interference.
In all figures, the input power corresponds to the launched power per subcarrier and the transmission distance is  $d=1000$ km.
 
\begin{figure}[h!]
	\centering		
	\includegraphics[width=.95\linewidth]{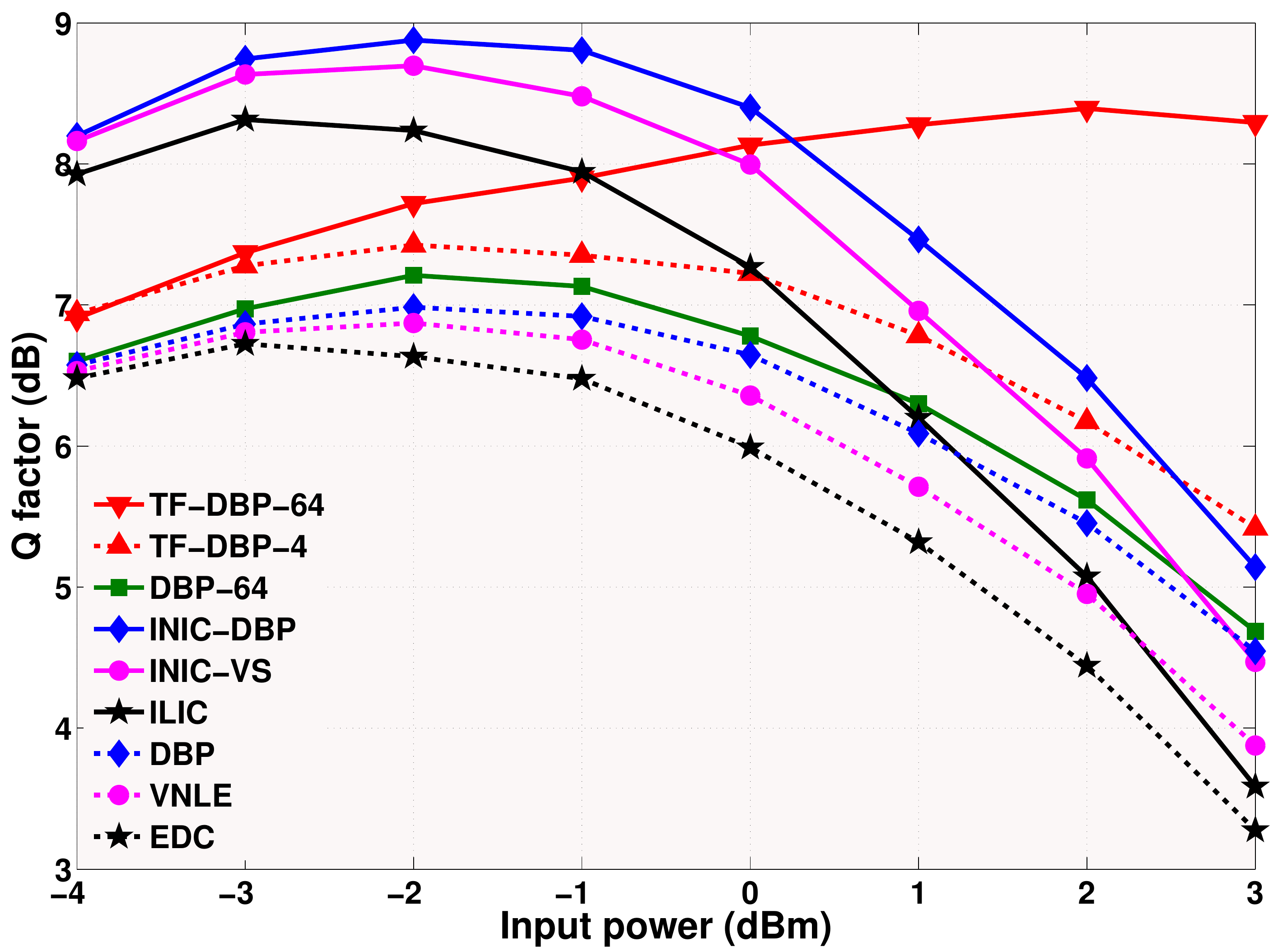}
%	\captionsetup{justification=centering}
	\caption{Q factor vs. the input power for $\rho=0.1$ and $\Delta = 1$. }	
	\label{fig:A}
\end{figure}

In Fig.~\ref{fig:A}, we show the Q factor versus the input power for an RRC roll-off factor $\rho=0.1$ and in the context of Nyquist WDM superchannel ($\Delta = 1$). We evaluate the performance of different techniques, which can be classified into three categories:
\begin{itemize}
\item Techniques applied per subcarrier: VNLE, single step DBP, DBP with 64 steps per span (DBP-64) and the linear electronic dispersion compensation (EDC). 
\item Techniques applied for the total field or all subcarriers: TF-DBP with 4 steps per span (TF-DBP-4) and TF-DBP with 64 steps per span (TF-DBP-64).
\item DFE-based techniques: INIC-DBP, INIC-VS and ILIC. These approaches are based on single step DBP, VNLE and EDC applied subcarrier per subcarrier.
\end{itemize}

As presented in Fig.~\ref{fig:A}, NLC techniques applied per subcarrier, like single step DBP and VNLE show limited performance because of the high impact of nonlinear and linear inter-subcarrier interference. In addition, the gain of DBP-64 is about $0.2$ dB in comparison with single step DBP. Therefore, the performance of DBP per subcarrier is still limited even when the number of steps per span is very high. 
On the other hand, TF-DBP-4 and TF-DBP-64 exhibit better performance in comparison with the NLC techniques applied per subcarrier. The gain of TF-DBP-64 and TF-DBP-4 is about $1.3$ dB and $0.5$ dB in comparison with DBP, respectively. TF-DBP-64 strongly outperforms TF-DBP-4. Therefore, because of the large bandwidth, TF-DBP requires a high number of steps to increase significantly the performance. TF-DBP-64 increases also the optimum input power, and then higher modulation formats can be used. 
The DFE-based approaches INIC-DBP and INIC-VS exhibit better performance than TF-DBP-64 and the gain is about $0.4$ dB and $0.2$ dB respectively. INIC-DBP and INIC-VS also strongly outperform single step DBP applied per subcarrier and the gain is about $2$ dB and $1.8$ dB, respectively. INIC-DBP and INIC-VS take into account both linear and nonlinear interference between subcarriers, while ILIC takes into account only the linear interference, which explain the gain of performance between them. 

At high input power, TF-DBP-64 shows better results than INIC approaches, which means that the TF-DBP manages the nonlinear interference better. In fact, INIC-DBP and INIC-VS take into account only a part of the nonlinear interference, as explained in Section \ref{sec-inic}. In addition, the reduced performance of the INIC approaches can be explained also by the fundamental limitation of the DFE. In fact, the INIC approaches use the detected symbols based on the NLC techniques applied subcarrier per subcarrier (first step of INIC). When the Q factor of the first step is very low (BER very high), the final decision (third step of INIC) can be affected by the error propagation, and then the Q factor of step three will be very low, as well.    
\begin{figure}[h!]
	\centering		
    \includegraphics[width=.95\linewidth]{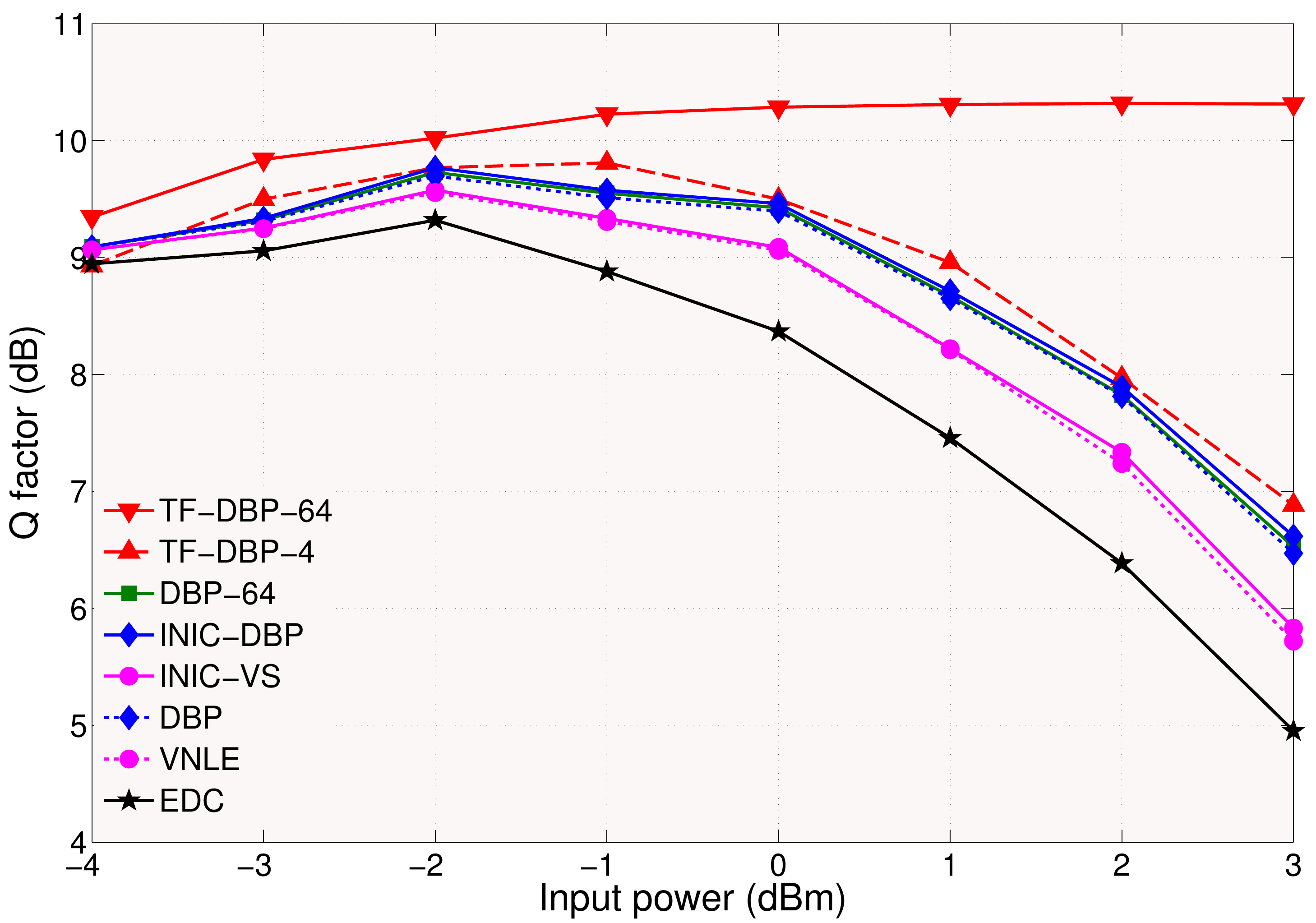}
	%\captionsetup{justification=centering}
	\caption{Q factor vs. the input power for $\rho=0.1$ and $\Delta = 1.1$.}	
	\label{fig:B}
\end{figure}

In Fig.~\ref{fig:B}, we plot the Q factor versus the input power for a subcarrier spacing factor $\Delta = 1.1$. In this case, there is no linear crosstalk between the optical subcarriers and only nonlinear interference exists. TF-DBP-64 shows the best performance in terms of the Q factor and nonlinear threshold. At $3$ dBm input power, the gain of TF-DBP-64  is about $3.4$ dB in comparison with TF-DBP-4. Therefore, significantly increasing the number of steps per span for TF-DBP leads to a significant increase of performance. On the other hand, the performances of INIC-DBP and INIC-VS are close to DBP and VNLE, respectively. This can be explained by the fact that the INIC techniques take into account only a part of the nonlinear interference because of the causality issue, as mentioned in Section \ref{sec-inic}. Note that, because of the absence of linear crosstalk, the performance of ILIC is not shown in this figure; it is exactly the same as the case of EDC.

\begin{figure}[h!]
	\centering		
    \includegraphics[width=.99\linewidth]{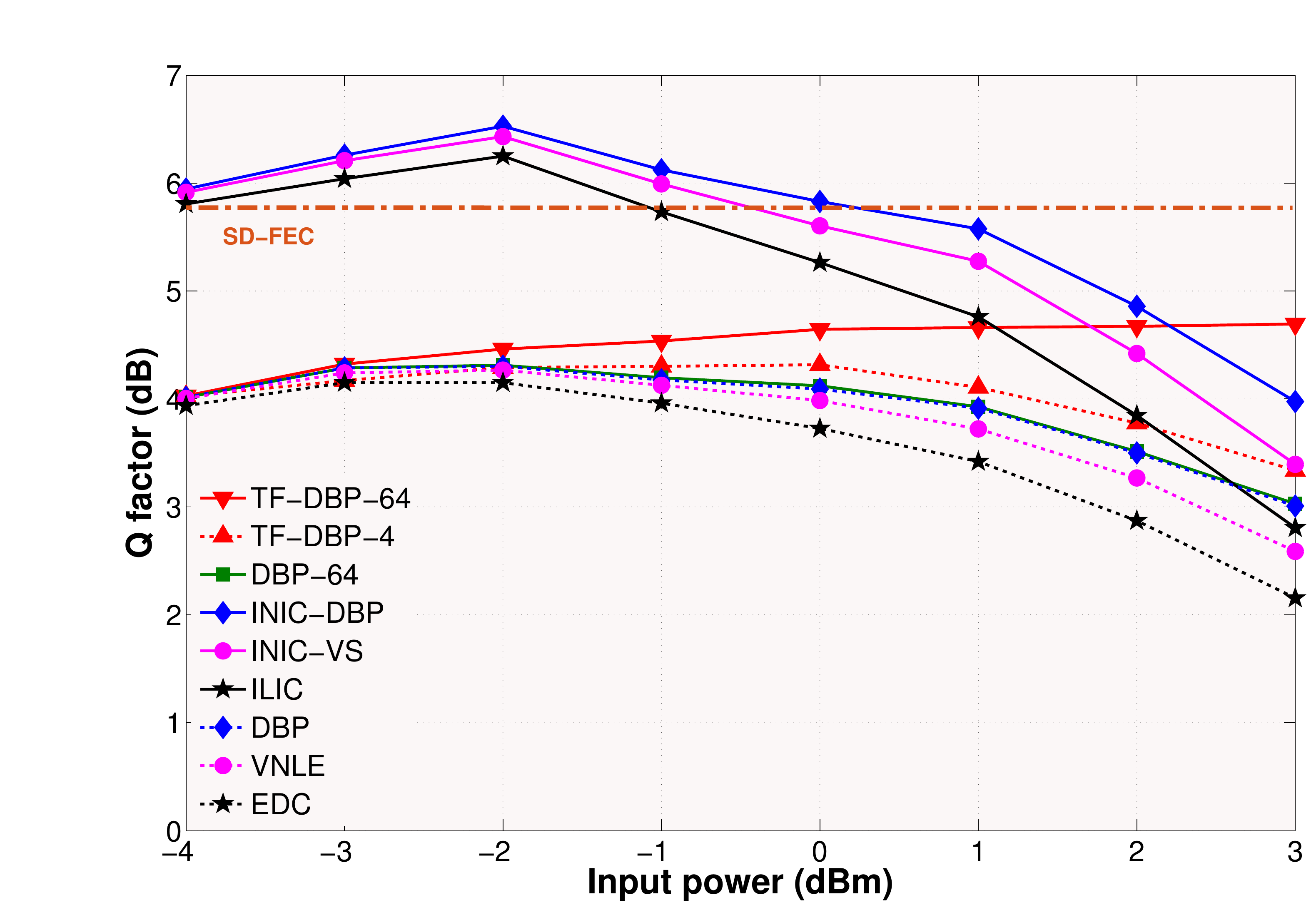}
	%\captionsetup{justification=centering}
	\caption{Q factor vs. the input power for $\rho=0.1$ and $\Delta = 0.95$.}	
	\label{fig:BB}
\end{figure}
In Fig.~\ref{fig:BB}, we plot the Q factor as a function of the input power for super-Nyquist WDM transmission ($\Delta = 0.95$). The three DFE-based techniques exhibit the best performance in comparison to other techniques. This is due to the cancellation of the linear crosstalk, which has a high impact in such a transmission system. INIC-DBP shows better performance in terms of the Q factor and nonlinear threshold. Compared to INIC-VS and ILIC, the gain in the nonlinear threshold is about $0.8$ dB and $1.4$ dB at soft-decision (SD)-FEC limit (Q = $5.9 dB$), respectively. The performance of TF-DBP techniques, in addition to VNLE and DBP applied per subcarrier, is strongly affected by the linear interference, and the Q factor is below the SD-FEC limit.

\begin{figure}[h!]
	\centering		
	\includegraphics[width=.95\linewidth]{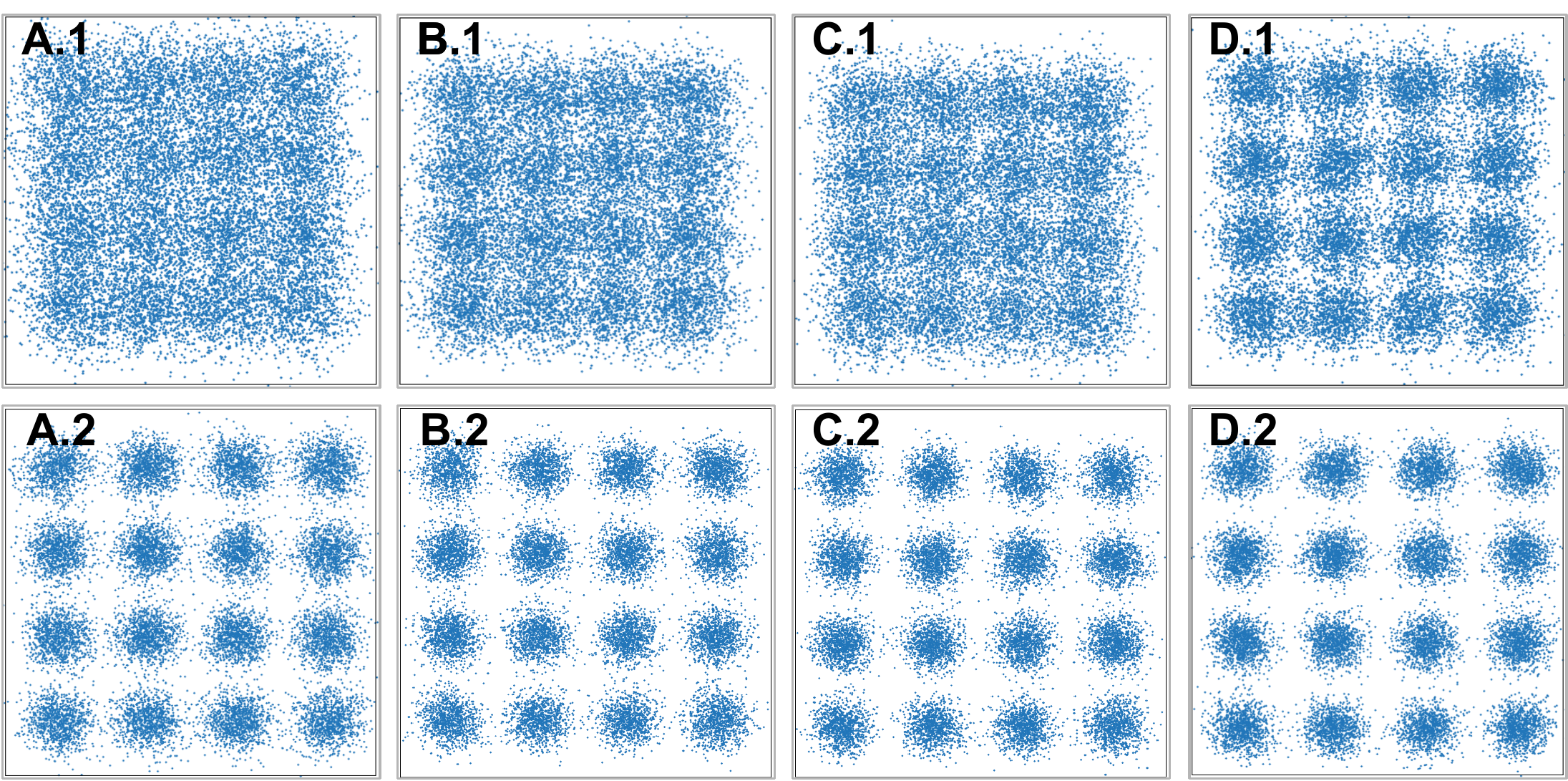}
%	\captionsetup{justification=centering}
	\caption{Constellation points for $\rho = 0.01$; 
	$\Delta = 0.95$:  A.1: EDC, B.1: DBP, C.1: TF-DBP-64, D.1: INIC-DBP; and 
	$\Delta = 1$:  A.2: EDC, B.2: DBP, C.2: TF-DBP-64, D.2: INIC-DBP.}	
	\label{fig:C}
\end{figure}

In Fig.~\ref{fig:C}, we focus on the DBP, TF-DBP and INIC-DBP techniques. We show the $16$QAM constellation points at optimum input power for an RCC roll-off factor $\rho = 0.01$ and different values of the subcarrier spacing factor ($\Delta = 0.95$ and $\Delta = 1$). Note that the optimum input power is the power providing the best performance in terms of the Q factor. 
When $\Delta = 0.95$, INIC-DBP outperforms TF-DBP-64, DBP and CDE. TF-DBP-64 and DBP present a similar performance to EDC because of the high impact of linear interference. For $\Delta = 1$, a comparable performance is observed for INIC-DBP and TF-DBP-64. The constellations after these two techniques are clearer and the points are slightly more visible than DBP, which would correspond to better results in terms of the Q factor.

\subsection{Complexity analysis} 
In this section, a complexity analysis is performed for the DBP, TF-DBP, VNLE, INIC-DBP, INIC-VS and EDC. The required number of real multiplications is used for the complexity evaluation. EDC requires $4 N_f \log_2(N_f)+ 4 N_f$ real multiplications \cite{Lui12}, where $N_f$ corresponds to the FFT size. The complexity of single-step DBP and VNLE are $C_{\textrm{DBP}} = 4 N N_f \log_2(N_f)+ 10.5 N N_f$ and $C_{\textrm{VNLE}}= 2 N N_f \log_2(N_f)+ 4.25 N N_f$, respectively, where $N$ is the number of spans. The INIC approach roughly triples the complexity because of the three steps implementation based on DFE. Note that the complexity of applying the IVSTF kernel $K_3$ is the same as the case of applying the VSTF kernel $H_3$. As in \cite{AA16}, we neglect the complexity of the extra DSP of step three of the INIC implementation, which can be initialized with the output of step one. Then, INIC-DBP and INIC-VS are three times more complex than DBP and VNLE, respectively. 

\begin{figure}[h!]
	\centering		
	\includegraphics[width=.95\linewidth]{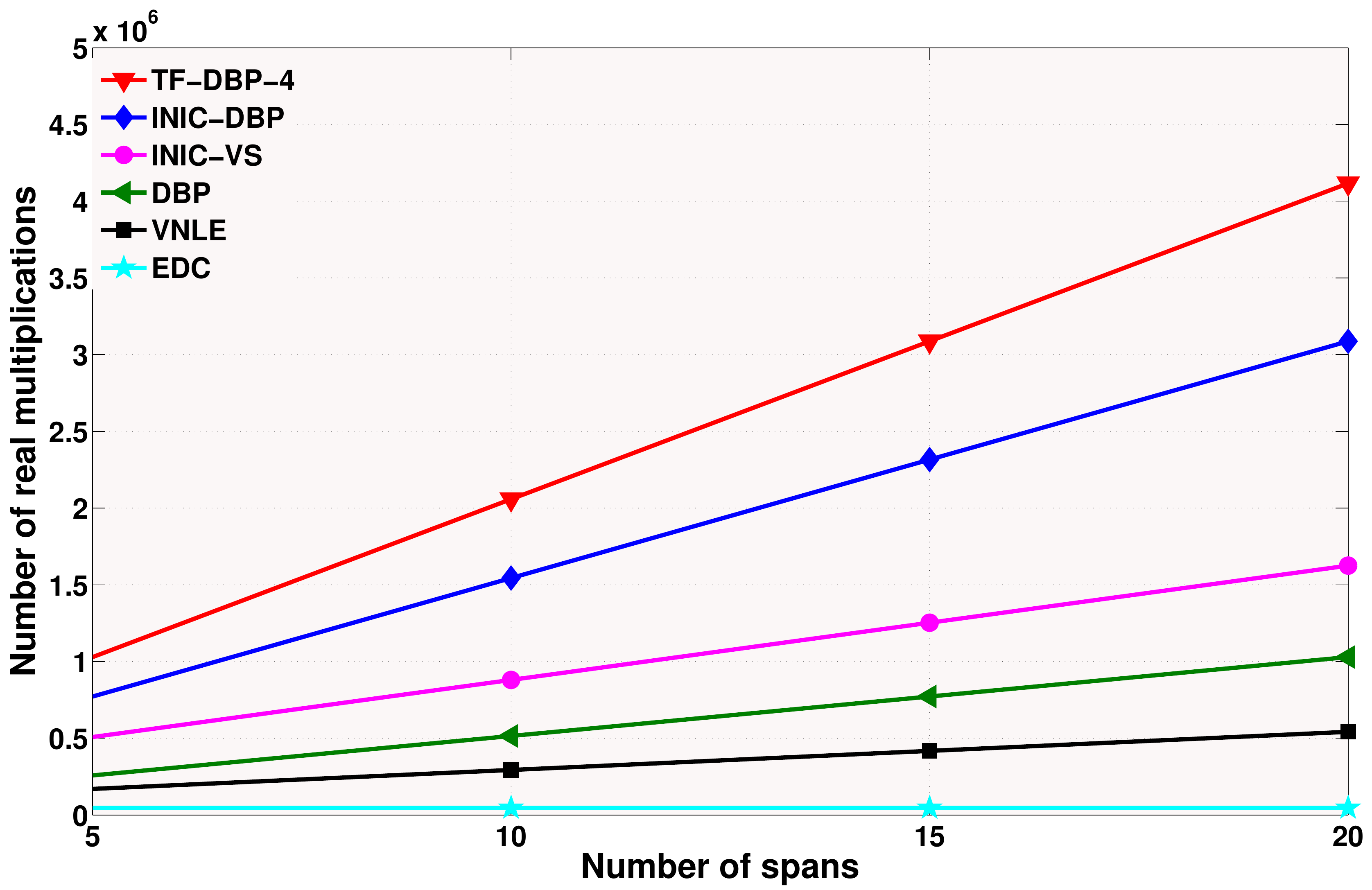}
%	\captionsetup{justification=centering}
    \vspace{-0.5cm}
	\caption{NLC complexity analysis.}	
	\label{fig:comp}
\end{figure}

Fig.\ref{fig:comp} shows the complexity of TF-DBP-4, INIC-DBP, INIC-VS, DBP, VNLE and EDC as a function of the number of spans. The FFT size used for complexity evaluation is $1024$.
INIC-based on single-step DBP still has relatively lower complexity in comparison with the TF-DBP with 4 steps per span.

 \vspace{-0.15cm}
\subsection*{Discussion:}
In terms of complexity and performance, INIC-DBP outperforms TF-DBP-64 when the subcarrier spacing factor is lower than or equal to 1 ($\Delta \leq 1$), as shown in Figs. \ref{fig:A} and \ref{fig:BB}, and it has lower complexity than TF-DBP-4. On the other hand, TF-DBP-64 outperforms INIC when the subcarrier spacing factor is higher than the symbol rate, according to Fig. \ref{fig:B}. In addition, this technique exhibits better results at high input power when compared with INIC-DBP, and then higher modulation formats can be used when applying TF-DBP. However, TF-DBP has high complexity and it faces the constraint of the unavailability of high speed DAC/ADC. Concerning the DBP and VNLE applied per subcarrier, these techniques exhibit limited performance, while they have lower complexity when compared with INIC and TF-DBP. 

It is worth mentioning that simulations are performed using the transmission parameters summarized in Table. \ref{tab:table1}. In general, increasing the bit rate by increasing the order of the modulation format or the symbol rate leads to the increase of the intra-subcarrier nonlinear effects, and thus, to the degradation of the transmission performance. The bit rate can also be increased by adding more subcarriers. In this case, inter-subcarrier nonlinear interference, such as XPM, XPolM and FWM, will increase. In such a configuration, the performance gap between the TF-DBP and INICs on one hand, and the DBP and VNLE per subcarrier on the other hand, will significantly increase. For ultra-long haul communications, in addition to the deterministic nonlinear effects, the accumulation of ASE noise and its interaction with the signal lead to a significant performance limitation.

\section{Conclusion and outlook} \label{section_4}
In this paper, we provided a comprehensive survey of fiber nonlinearity compensation techniques. We started with a brief description of the optical link nonlinear effects; these effects increase with the data rate and are inversely proportional to the channel/subcarrier spacing. High data rate and reduced subcarrier spacing characterize next generation WDM communication systems, which result in a strong reduction of the transmission performance of such systems due to the fiber nonlinear effects. Following this overview of nonlinear impairments, several NLC techniques were presented with a focus on the promising approaches. In addition to the principle of these techniques, a highlight of their advantages and drawbacks in terms of complexity, hardware requirements and performance were presented to ensure that an interested reader is provided with a general comparison of the NLC techniques. NLC techniques, such as DBP and VNLE significantly improve the transmission performance. However, such approaches are complex for real implementation and their performance is affected by nonlinear interference in superchannel systems. Techniques taking into account the inter-subcarrier interference, like INIC, NFT and TF-DBP improve the performance in superchannel systems, but they increase the complexity of implementation as well. On the other hand, PC techniques exhibit low complexity in comparison with DBP and VNLE, while OPC affects the flexibility of the optical network and PCTW halves the spectral efficiency. The perturbation-based NLC can be implemented in a single stage for the entire link and with one sample per symbol, which reduces the hardware requirement. However, this technique employs a large number of perturbation terms.  

Furthermore, we evaluated the performance of the well-known NLC techniques and the proposed INIC-DBP in the context of Nyquist and super-Nyquist WDM superchannel. In addition, a complexity analysis of these techniques was provided, so that a compromise between performance and complexity can be seen. NLC techniques applied per subcarrier like, DBP and VNLE exhibit very limited performance in the context of Nyquist and super-Nyquist systems. INIC and TF-DBP approaches significantly increase the performance. TF-DBP present better performance at high input power while INICs are more suitable to super-Nyquist systems than TF-DBP. In terms of complexity, INIC-DBP has ower complexity in comparison the TF-DBP. 
 
For future works in this research area, three main research paths are open:
\begin{itemize}
\item Complexity reduction
\end{itemize}
 NLC is a cost effective key technology to increase the data rate in the next generation WDM communication systems. However, the main challenges for the commercial use of the NLC techniques is the complexity of implementation and flexibility. The near future work should be in the direction of proposing low complex and flexible NLC approaches to be commercially implemented. This can be done by reducing the complexity of the existing techniques or by finding new low complexity approaches without loss in performance. 
 \begin{itemize}
\item Performance improvement
\end{itemize}
The majority of the proposed NLC techniques have focused on the mitigation of nonlinear deterministic effects without considering the interaction of the transmitted signal with the ASE noise. Such interaction can also be a strong limitation of the transmission performance, especially in case of very long transmission distance. In addition, the majority of NLC approaches does not consider the interaction between the nonlinear effects and the non-deterministic linear effects, such as PDL and PMD. A study of such interactions should also be carried out.  

DFE-based NLC techniques dealing with the digital information, such as the proposed INIC-DBP, can be extended to compensate for the intra-subcarrier nonlinear effects and all inter-subcarrier nonlinear interference. That can lead to an improved transmission performance.  

More studies should be conducted also for the NFT-based communication, which represents a promising approach to handle the fiber nonlinearity. As, NFT-based communication is not affected by all deterministic linear and nonlinear effects including intra-subcarrier and inter-subcarrier cross-talk, it can significantly increase the transmission performance.   
 \begin{itemize}
\item Future systems
\end{itemize}
The so-called "capacity crunch" due to the full exploitation of the installed network resources, pushes the researchers to focus on other technological paths to
increase the network capacity in the long-term future. The SDM techniques appear to be the most promising alternative paths to increase the optical transmission capacity. The SDM techniques consist in increasing the capacity by using multi-mode or multi-core fibers instead of SMF. Other approaches, such as the use of the (C+L) EDFA band transmission or the hybrid EDFA-Raman amplification can be a solution to increase the capacity, as well.

\vspace{0.35cm}
\textbf{\textit{List of acronyms and abbreviations}}
\vspace{0.15cm}

\begin{abbrv}
\small
\item[ADC]  Analog-to-digital converter
\item[AWGN] Additive white Gaussian noise
\item[ASE] Amplified spontaneous emission 
\item[BER]  Bit error rate
\item[CD]  Chromatic dispersion
\item[DAC]  Digital-to-analog converter
\item[DBP]  Digital back-propagation
\item[DFE]  Decision feedback equalizer
\item[DP]  Dual-polarization
\item[DSP] Digital signal processing 
\item[EDC] Electronic dispersion compensation
\item[EDFA] Erbuim-doped fiber amplifier
\item[FEC] Forward error correction 
\item[FFT] Fast Fourier transform
\item[FWM] Four wave mixing
\item[IFFT] Inverse fast Fourier transform
\item[INIC] Inter-subcarrier nonlinear interference canceler
\item[ILIC] Inter-subcarrier linear interference canceler
\item[IVSTF] Inverse Volterra series transfer function
\item[NLSE] Nonlinear Schr\"odinger equation
\item[NLC] Nonlinearity compensation
\item[OFDM] Orthogonal frequency division multiplexing
\item[OPC] Optical phase conjugation
\item[OSNR] Optical signal-to-noise ratio
\item[PC] Phase conjugation
\item[PCTW] Phase conjugated twin waves
\item[PDL] Polarization dependent loss
\item[PMD] Polarization mode dispersion
\item[QAM] Quadrature amplitude modulation
\item[QPSK] Quaternary phase shift keying
\item[RRC] Root raised-cosine 
\item[SBS] Stimulated Brillouin scattering
\item[SDM] Space division multiplexing
\item[SOP] State of polarization
\item[SPM] Self-phase modulation
\item[SMF] Single mode fiber
\item[SRS] Stimulated Raman scattering
\item[SSFM] Split-step Fourier method 
\item[VSTF] Volterra series transfer function
\item[VNLE] Volterra nonlinear equalizer
\item[VNI] Visual networking index
\item[WDM] Wavelength division multiplexing
\item[W-VSNE] Weighted Volterra series nonlinear equalizer
\item[XPM] Cross-phase modulation
\item[XPolM] Cross-polarization modulation
\end{abbrv}
\bibliographystyle{IEEEtran}
\bibliography{Biblio}
\end{document}